\documentclass{article}

\usepackage{microtype}
\usepackage{graphicx}
\usepackage{subcaption}
\usepackage{booktabs} 

\usepackage{hyperref}
\usepackage{pifont}

\usepackage[accepted]{icml2026}

\usepackage{bbding} 
\usepackage{amsmath}
\usepackage{amssymb}
\usepackage{mathtools}
\usepackage{amsthm}
\usepackage{graphicx}
\usepackage{subcaption}
\usepackage{tcolorbox} 
\usepackage{multirow}
\usepackage{enumitem}
\usepackage{booktabs}
\usepackage{xspace}
\usepackage{array}
\usepackage{makecell}
\usepackage[capitalize,noabbrev]{cleveref}

\theoremstyle{plain}

\theoremstyle{definition}

\theoremstyle{remark}

\newcommand{\alias}{\texttt{TRAP}} 

\newcommand{\hzx}[1]{{\color{black}{#1}}}
\usepackage[textsize=tiny]{todonotes}

\icmltitlerunning{\alias: Hijacking VLA CoT-Reasoning via Adversarial Patches}

\begin{document}
\setlength{\abovedisplayskip}{6pt}
\setlength{\belowdisplayskip}{6pt}
\twocolumn[
  \icmltitle{\alias: Hijacking VLA CoT-Reasoning via Adversarial Patches}



  \icmlsetsymbol{equal}{*}
  \icmlsetsymbol{cor}{\Envelope}

  \begin{icmlauthorlist}
    \icmlauthor{Zhengxian Huang}{yyy}
    \icmlauthor{Wenjun Zhu}{yyy}
    \icmlauthor{Haoxuan Qiu}{comp}
    \icmlauthor{Xiaoyu Ji}{yyy,cor}
    \icmlauthor{Wenyuan Xu}{yyy}
  \end{icmlauthorlist}

  \icmlaffiliation{yyy}{Zhejiang University, Hangzhou, China}
  \icmlaffiliation{comp}{Harbin Institute of Technology, Harbin, China}

  \icmlcorrespondingauthor{Xiaoyu Ji}{xji@zju.edu.cn}

  \icmlkeywords{Machine Learning, ICML}

  \vskip 0.3in
]



\printAffiliationsAndNotice{}  

\begin{abstract}
By integrating Chain-of-Thought (CoT) reasoning, Vision-Language-Action (VLA) models have demonstrated strong capabilities in robotic manipulation, particularly by improving generalization and interpretability. However, the security of CoT-based reasoning mechanisms remains largely unexplored. In this paper, we show that CoT reasoning introduces a novel attack vector for targeted behavior hijacking—for example, causing a robot to mistakenly deliver a knife to a person instead of an apple—without modifying the user’s instruction. We first provide empirical evidence that CoT strongly governs action generation, even when it is semantically misaligned with the input instructions. Building on this observation, we propose \textbf{TRAP}, the first targeted behavior-hijacking adversarial attack against CoT-reasoning VLA models. By targeting the reasoning-to-action pathway, \textbf{TRAP} uses an adversarial patch (\textit{e.g.}, a tablecloth placed on the table) to steer intermediate CoT reasoning and downstream actions toward adversary-defined behaviors. Extensive evaluations on three representative reasoning VLAs, spanning distinct CoT reasoning mechanisms, demonstrate the effectiveness of \textbf{TRAP}. Notably, we implemented the patch by printing it on paper in a real-world setting. Our findings highlight the urgent need to secure CoT reasoning in VLA systems. The project page is available at \href{https://zhengxian-huang.github.io/TRAP-website/}{TRAP-website}.

\end{abstract}

\section{Introduction}

Vision–Language–Action (VLA) models~\cite{kim2024openvla} have reshaped robot learning by enabling open-world manipulation through end-to-end training on large-scale robot data. Recently, a growing body of work has \hzx{explored} reasoning VLAs~\cite{zawalski2024robotic,black2025pi_}, which incorporate Chain-of-Thought (CoT) reasoning to improve generalization to new scenes and tasks. Beyond performance, explicit reasoning capabilities are also argued to enhance interpretability and safety~\cite{zawalski2024robotic} by involving intermediate decision-making steps.

Although CoT-enabled reasoning VLA models have attracted growing attention, their security remains largely unexplored. Existing studies primarily examine adversarial attacks on vanilla VLA systems by perturbing perception~\cite{lu2025robots, yan2025alignment} or disrupting action generation~\cite{wang2025exploring, jones2025adversarial}. Importantly, most prior attacks are untargeted, leading to generic performance degradation. 
Different from existing attack vectors, CoT often makes the model’s task intent explicit, effectively leaking high-level goals and intermediate plans. \hzx{In manipulation tasks, CoT can encode intermediate intent as target bounding boxes, motion sequences, or trajectories.} This motivates a central question of this work: Can an attacker hijack a VLA model’s \hzx{behavior} by manipulating its CoT reasoning?

In this paper, we show that CoT reasoning can indeed be used for targeted \hzx{behavior} hijacking—for example, causing a robot to mistakenly deliver a knife to a person instead of an apple—without modifying the user’s instruction, as illustrated in the threat scenario of Fig.~\ref{fig:method}. To investigate how CoT reasoning influences the action generation of VLA models, we conducted a preliminary experiment by actively intervening in instruction–CoT pairs. Our empirical results reveal that CoT strongly governs action generation, even when it is semantically misaligned with the input instructions.
Motivated by this insight, we propose \textbf{TRAP} (Co\textbf{T}-\textbf{R}easoning \textbf{A}dversarial \textbf{P}atch), the first targeted \hzx{behavior-hijacking adversarial} attack for reasoning VLAs. \textbf{TRAP} uses an adversarial patch (\textit{e.g.}, a \hzx{tablecloth} placed on the table) to corrupt intermediate CoT reasoning and hijack the VLA’s output. Unlike attacks based solely on action losses,
\textbf{TRAP} induces specific and adversary-defined behaviors by optimizing the CoT adversarial loss. 
\hzx{We validate \textbf{TRAP} on three VLA architectures, each adopting a distinct CoT reasoning paradigm and collectively covering both integrated and hierarchical CoT mechanisms, demonstrating the broad applicability of our attack.}
\hzx{Notably, we demonstrate real-world attacks with visually plausible printed patches, enabled by stealthiness-aware optimization, homography transformation, and color calibration.} Our findings highlight the urgent need to secure CoT reasoning in VLA systems.

Our main contributions are summarized as follows:
\begin{itemize}[topsep=-1pt,parsep=-1pt,partopsep=-1pt]
    \item \textbf{New attack \hzx{vector} discovery.} We identify a novel attack vector in CoT reasoning VLAs, where the introduction of CoT mechanisms expands the attack surface and makes it possible to manipulate the behavior of VLAs via adversarial patches.
    \item \hzx{\textbf{Targeted behavior-hijacking attack against reasoning VLAs.}
    We propose \textbf{TRAP}, a targeted adversarial attack framework that optimizes a physical patch to hijack intermediate CoT reasoning and steer VLA actions toward attacker-specified behaviors without modifying the user's instruction.
    To the best of our knowledge, this is the first adversarial attack that exploits CoT reasoning to induce targeted behavior hijacking in VLAs.}
    \item \hzx{\textbf{Extensive evaluation across reasoning VLA paradigms.}
    We evaluate \textbf{TRAP} on three representative reasoning VLAs, spanning diverse CoT reasoning mechanisms and model architectures.
    Results from simulation and real-world physical deployments demonstrate the effectiveness and broad applicability of our attack.}
\end{itemize}

\section{Background and Related Work}
\subsection{Reasoning Vision-Language-Action Models}
\label{sec:reasoning_vla}
Vision-Language-Action Models (VLAs)~\cite{zitkovich2023rt,kim2024openvla} are multi-modal models that can process visual and instruction inputs and generate robot actions. Trained on a large-scale robotic dataset~\cite{o2024open}, VLAs show promising performance on general daily-life tasks~\cite{pi0_2024}. Despite the significant progress in VLAs, they still struggle with limited generalization as well as limited performance on long-horizon tasks. 

\textbf{CoT-reasoning VLAs.} Inspired by the success of Chain-of-Thought (CoT) reasoning in LLMs, recent VLA models have incorporated embodied CoT mechanisms, a paradigm often referred to as ``reasoning VLAs". Specifically, after perceiving visual and instruction inputs, the reasoning VLA first generates a CoT of diverse forms, such as (1) subtask decomposition~\cite{pi0_2024,chen2025internvla,yang2025instructvla}, (2) object bounding boxes~\cite{deng2025graspvla}, or (3) predicted traces~\cite{lee2025molmoact}. Subsequently, the final action is generated by conditioning both the CoT and the inputs. Thus, the generation of CoT-reasoning VLAs can be formulated as:
\begin{gather}
    r_t \sim P_\theta(r_t \mid r_{<t}, O, I)\\
    \quad a \sim P_\theta(a \mid R, O, I)
\end{gather}
where $P_\theta$ denotes the reasoning VLA parameterized by $\theta$. $O$ represents the environmental observations (including visual inputs), and $I$ is the user instruction. Furthermore, $r_t$ denotes the $t$-th token of the generated CoT $R$, while $a$ corresponds to the predicted robot action.

\begin{figure}[t]
  \begin{center}
    \includegraphics[width=0.9\columnwidth]{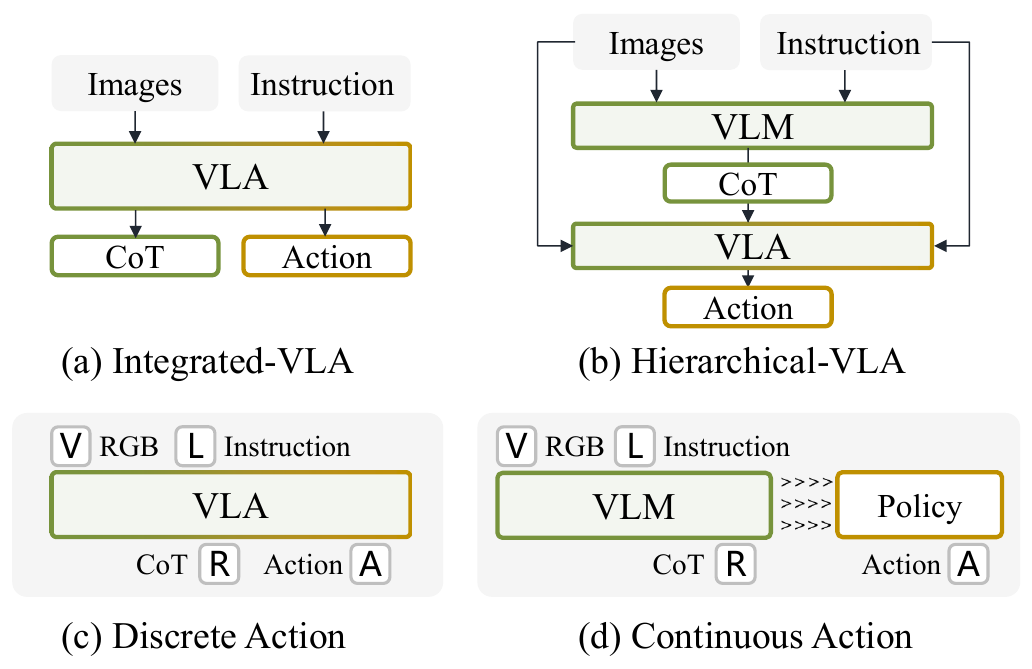}
    \caption{Illustration of existing representative paradigms of reasoning VLAs.}
    \label{fig:reasoning VLA}
  \end{center}
  \vspace{-15pt}
\end{figure}

CoT-reasoning VLAs can be broadly categorized into Integrated-VLAs and Hierarchical-VLAs~\cite{gao2026vla}, as illustrated in Figure~\ref{fig:reasoning VLA}. The Integrated-VLA is a unified model that jointly performs reasoning and action generation. It can be further divided into discrete-token methods~\cite{zawalski2024robotic,sun2025emma}, \hzx{such as MolmoAct~\cite{lee2025molmoact}}, which quantize actions into discrete bins and model them as tokens via autoregressive generation, and continuous-regression methods~\cite{black2025pi_,yin2025deepthinkvla}, \hzx{such as GraspVLA~\cite{deng2025graspvla}}, which use diffusion models, flow matching, or MLP heads to directly regress continuous actions. In contrast, the Hierarchical-VLA~\cite{chen2025internvla,shi2025hi}, \hzx{such as InstructVLA~\cite{yang2025instructvla}}, typically adopts a dual-system design: a VLM serves as a high-level planner that generates CoT, which then guides a low-level policy (\textit{e.g.}, a vanilla VLA) to execute precise actions.

CoT mechanisms can improve the performance of \hzx{VLAs}, but they also introduce a new attack vector. We present the first evidence that an attacker can hijack the CoT reasoning process to manipulate a VLA’s downstream actions and induce targeted behaviors. We further show that this vulnerability is general, persisting across diverse CoT paradigms and VLA architectures, including both discrete and continuous action frameworks.

\subsection{Adversarial Attacks against VLAs}

Adversarial attacks against VLAs have attracted significant attention, particularly because compromised VLAs may cause catastrophic physical consequences in real-world environments. \hzx{Prior work has primarily focused on \textbf{untargeted attacks}, which introduce adversarial perturbations into visual or language inputs to induce task failures.} Existing untargeted attacks include (1) perception-based approaches~\cite{lu2025robots,yan2025alignment,xu2025model}, which undermine scene understanding and cross-modal alignment, and (2) action-based approaches~\cite{wang2025exploring,jones2025adversarial,wang2025freezevla}, which directly disrupt action generation. For example, RoboticAttack~\cite{wang2025exploring} optimizes adversarial patches using action-guided loss objectives, whereas UPA-RFAS~\cite{lu2025robots} attacks visual-encoder representations via a unified feature-space objective coupled with attention steering.

However, targeted \hzx{behavior-hijacking adversarial} attacks against VLAs remain largely underexplored, despite posing a substantially greater threat by inducing targeted malicious behaviors. Moreover, existing studies have primarily focused on vanilla VLA architectures, leaving the vulnerabilities of emerging CoT-reasoning VLAs largely \hzx{overlooked}. In this paper, we bridge this gap by analyzing the security of CoT-reasoning VLAs and designing targeted attacks that explicitly manipulate \hzx{VLAs' behavior}.

\section{Threat Model}
\begin{figure*}[t]
  \begin{center}
    \includegraphics[width=0.98\textwidth]{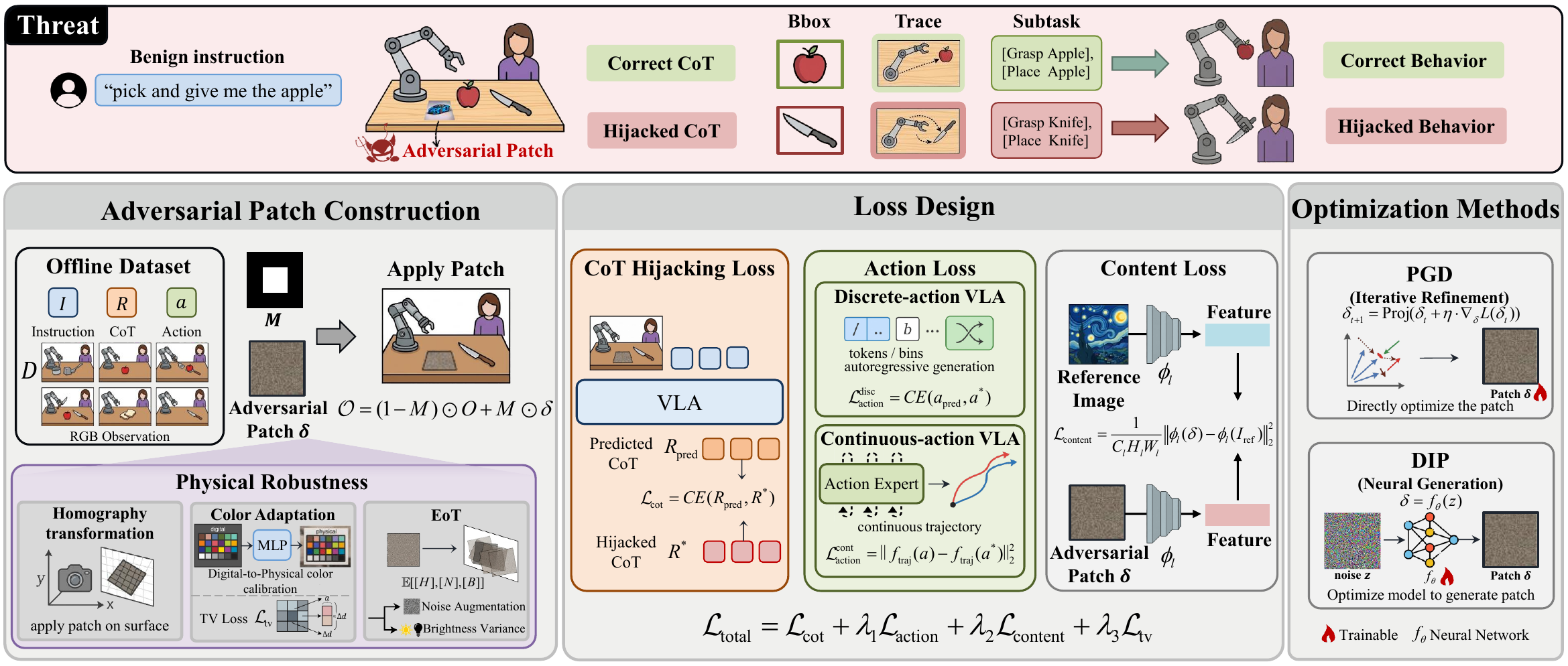}
    \caption{
    \textbf{Overview of the TRAP attack framework.}
    The adversary places an adversarial patch in the scene to corrupt the VLA's intermediate CoT, causing the model to execute an attacker-specified behavior while the user instruction remains benign.
    The patch is optimized with two groups of objectives: attack-effectiveness losses, including CoT hijacking and action losses, and stealthiness losses, including content and TV losses.
    }
    \label{fig:method}
  \end{center}
  \vspace{-15pt}
\end{figure*}

\textbf{Attack Goal.} The attacker’s primary objective is to induce a VLA to execute a \emph{targeted} malicious behavior (\textit{e.g.}, handing a knife to the user), as illustrated in Figure~\ref{fig:method}. Concretely, the attacker can introduce a printed adversarial patch into the scene (\textit{e.g.}, a tablecloth on the table), which biases the model’s CoT reasoning and consequently causes malicious or unintended behavior. \hzx{The attack is further designed to persist across multiple VLA inference steps, reflecting the short-horizon nature of current VLAs, which typically predict only one or a few actions per inference.}

\textbf{Adversary Capability.} Following prior work~\cite{wang2025exploring}, we consider a white-box adversary who has access to the victim VLA’s architecture, parameters, and gradients. In the black-box setting, the attacker can transfer an adversarial patch optimized on a surrogate VLA. We further assume the adversary can physically deploy the patch in the environment (\textit{e.g.}, by sticking the patch on the table). \hzx{For reliable physical realization, the adversary may obtain deployment-specific calibration priors, such as homography estimation and a color calibration model, but these priors do not grant any control beyond physical patch placement.} Importantly, the adversary cannot manipulate the instruction input, \textit{i.e.}, all instructions remain benign user-provided commands.

\section{Preliminary Analysis}
\label{sec:prelimiary}
Despite the widespread adoption of CoT in recent VLA models~\cite{deng2025graspvla,chen2025internvla,black2025pi_}, its underlying mechanisms, particularly its security implications, remain underexplored. ECoT-lite~\cite{chen2025training} suggests that ECoT’s gains primarily stem from improved representation learning induced by reasoning supervision, \hzx{and} VLA-OS~\cite{gao2026vla} provides a comprehensive empirical study of different \hzx{CoT} paradigms and their representational modalities. However, existing work has not disentangled the causal roles of instructions and CoT, either individually or jointly, in driving action generation. More critically, \hzx{when instructions and CoT conflict, the internal arbitration dynamics remain unclear}, leaving open which signal dominates the final action. These gaps motivate two key questions for enabling targeted \hzx{behavior-hijacking} attacks:
\begin{itemize}
    \item \textbf{Role of CoT in action generation}: \hzx{How much causal influence does CoT exert on downstream action generation, and can manipulating CoT reliably steer the model toward targeted actions?}
    \item \textbf{Conflict arbitration}: When input instructions and CoT are misaligned, does the model adhere more faithfully to the original instruction or to the intermediate reasoning?
\end{itemize}
To answer these two questions, we conduct a preliminary analysis of the causal mechanisms in reasoning VLAs. Specifically, we perform active interventions on the inputs of the VLAs and observe the subsequent action generation. We first collect original tuples $(I,R)$ and then consider two experimental settings:

\textbf{(1) Instruction Masking.} We assess the impact of CoT $R$ on action generation by masking instruction tokens, $P_\theta(a|R,O,\cdot)$. \hzx{This setting enables us to examine whether VLAs can complete tasks when provided with only CoT.}

\textbf{(2) Cross-Sample Shuffling.} To create semantic conflict, we disrupt the correspondence between the instruction and the CoT. For the $i$-th tuple, we replace its instruction $I^{(i)}$ with a distractor $I^{(j)}$ from a different task instance $j \neq i$ (\textit{e.g.}, $I^{(i)}$ is ``pick apple", and its corresponding CoT $R^{i}$ is the intermediate reasoning information for reaching the apple, while $I^{(j)}$ is ``pick knife". The action is generated via $P_\theta(a|R^{(i)}, O,I^{(j)})$. This setting creates conflict where $R^{(i)}$ and $I^{(j)}$ suggest different follow-up action patterns, allowing us to investigate which one dominates, and it is closer to the settings of our subsequent attack.

\textbf{Results and Analysis.} We examine three mainstream reasoning VLAs across various tasks to evaluate how their behaviors change under the two intervention settings relative to the original configuration (See Section~\ref{sec:exp_set} for full details on the experimental environment and metrics.). As shown in Table \ref{tab:preliminaries}, we observe the task performance degradation under the two aforementioned settings compared to $\text{TSR}_{\text{ori}}$, suggesting that instructions and CoT jointly contribute positively to action generation.
Under the instruction masking setting, the VLAs retain partial capability when driven solely by CoT, highlighting CoT’s positive influence on action generation.
Under the cross-sample shuffling setting, $\text{TSR}_{\text{s}^{i}}$ shows that CoT plays a more dominant role than the instruction, whereas $\text{TSR}_{\text{s}^{j}}$ suggests the opposite. 
Notably, the near-zero scores for both InstructVLA and MolmoAct imply that, across tasks, instruction and CoT act as competing signals with roughly comparable influence. \hzx{In contrast, the high score achieved by GraspVLA suggests that, for this model, CoT plays a dominant role in governing action generation.} Additional experimental details and analysis are provided in Section~\ref{append:pre}.
\begin{table}[t]
  \caption{\textbf{Instruction masking and CoT shuffling analysis on reasoning VLAs.} TSR denotes Task Success Rate. Subscripts $m, s^i, s^j$ denote mask and shuffling settings respectively.}
  \vspace{-10pt}
  \label{tab:preliminaries}
  \begin{center}
    \begin{small}
      \resizebox{\columnwidth}{!}{
        \setlength{\tabcolsep}{3pt}
        \begin{tabular}{lcccccc}
          \toprule
          Model & $\text{TSR}_{\text{ori}}$ & $\text{TSR}_{m}$ & $\text{Sim}_{m}$ & $\text{TSR}_{s^{i}}$ & $\text{TSR}_{s^{j}}$ & Score \\
          \midrule
        InstructVLA & 45.23\% & 39.95\% & 0.3169 & 16.42\% & 13.43\% & 0.0508 $\pm$ 0.2392\\
          MolmoAct    & 50.40\% & 17.35\% & 0.1805 & 12.94\% & 16.26\% & 0.0520 $\pm$ 0.1964\\
          GraspVLA    & 94.20\% & 92.60\% & 0.3990  & 94.20\%  &  0.00\% & 0.5278 $\pm$ 0.1696\\
          \bottomrule
        \end{tabular}
      }
    \end{small}
  \end{center}
  \vspace{-15pt}
\end{table}

\begin{tcolorbox} \noindent{\bf Key Observation:} \textit{CoT-reasoning VLAs exhibit a ``competition mechanism'' in which input instructions and CoT competitively influence action generation, with CoT acting as a vital driver of the final action.} \end{tcolorbox}

\section{Methodology}

Motivated by the pivotal role of CoT in reasoning VLAs in Section~\ref{sec:prelimiary}, we propose an adversarial patch attack designed to manipulate VLAs' behavior via CoT hijacking. In the following sections, we formalize the problem statement and detail our proposed attack.

\subsection{Problem Formulation}
\label{sec:problem}
\textbf{Adversarial Patch Injection. }The adversary aims to manipulate the VLAs to execute a target behavior by applying an adversarial patch to the environment. Let $\delta$ denote the adversarial patch and $M$ be a binary mask indicating the patch's position. For every time step $t$, the adversarial observation $\tilde{O}$ is formulated as:
\begin{equation} \tilde{O} = (1 - M) \odot O + M \odot \delta 
\end{equation}
where $\odot$ denotes element-wise multiplication. The adversarial patch should remain effective throughout the entire rollout.
\textbf{Offline Rollout. }Unlike previous patch attack work on a single static frame, our goal is to hijack the VLAs to execute a coherent, sequential target behavior. Thus, we curate an offline clean dataset $\mathcal{D}=\{\tau_n\}_{n=1}^N$ by rolling out the VLAs on different tasks. The trajectory $\tau$ is represented as a tuple $(O, R, a)$.

\textbf{Optimization Objective.} The adversarial objective is to optimize the patch $\delta$ such that when it is applied to the clean observations in $\mathcal{D}$, the VLA $P_\theta$ is induced to generate the target reasoning CoT $R^*$ and action $a^*$. \hzx{Additionally, to improve the stealthiness  of the patch, we further introduce a content loss $\mathcal{L}_{\text{content}}$ and Total Variation (TV) loss $\mathcal{L}_{\text{tv}}$.} The coefficients $\lambda_1$, $\lambda_2$, and $\lambda_3$ balance the contributions of the action loss, content loss, and TV loss, respectively. Formally, we minimize:

\begin{equation}
    \min_{\delta \in \Delta} \mathbb{E}_{\tau \sim \mathcal{D}} [\mathcal{L}_{\text{cot}}  + \lambda_1 \mathcal{L}_{\text{action}} +\lambda_2\mathcal{L}_{\text{content}}+\lambda_3\mathcal{L}_{\text{tv}}]
\end{equation}
\subsection{Attack Effectiveness}
\hzx{\textbf{CoT Hijacking Loss.}} \hzx{Prevailing reasoning VLAs predominantly employ VLMs to synthesize CoT reasoning. Despite variations in the specific form of CoT, its generation is commonly formulated as a standard next-token prediction task.} Thus, we employ a Cross-Entropy (CE) loss to align the generated CoT tokens with the target sequence $R^*$.
\begin{equation}
\mathcal{L}_{\text{cot}} = -\sum_{t=1}^{T} \log P_\theta(r^*_t \mid r^*_{<t}, \tilde{O}, I)
\end{equation}
\hzx{\textbf{Action Loss.}} As discussed in Section~\ref{sec:reasoning_vla}, existing VLA architectures utilize a variety of action-decoding heads—typically discrete tokenization or continuous regression. Therefore, we apply different action loss strategies. For VLAs that treat actions as discrete tokens, we utilize the CE loss:
\begin{equation}
    \mathcal{L}_{\text{action}}^{\text{disc}} = -\log P_\theta(a^* \mid R^*,\tilde{O}, I)
\end{equation}
For VLAs that regress to continuous action (\textit{e.g.}, diffusion-based), especially those employing action chunking~\cite{zhao2023learning}, we transform the action sequence into waypoints. We employ the Mean Squared Error (MSE):
\begin{equation}
    \mathcal{L}^{\text{cont}}_{\text{action}}= \| \text{f}_{\text{traj}}(a) - \text{f}_{\text{traj}}(a^*) \|^2_2
\end{equation}
where $\text{f}_{\text{traj}}(\cdot)$ maps the raw model output to the trajectory.

To optimize the adversarial patch $\delta$, we adopt Projected Gradient Descent (PGD)~\cite{madry2017towards} and the optimization process can be formulated as:
\begin{equation}
    \delta_{t+1}=\mathrm{Proj}_{||\cdot||_{\infty\leq\epsilon}}(\delta_t+\eta\cdot\nabla_\delta L(\delta_t))
\end{equation}
where $\mathrm{Proj}(\cdot)$ projects $\delta$ into perturbation budget $\epsilon$, $\eta$ is the attack step size, and the $\nabla L(\cdot)$ is the gradient of total loss.
\subsection{Stealthiness Enhancement}
\label{sec:stealthiness}
\hzx{\textbf{Content Loss.} Without camouflage constraints, adversarial patch optimization tends to produce semantically meaningless patterns that are visually conspicuous to humans. To improve the stealthiness of the attack, we follow~\cite{zhu2023tpatch} and introduce a content loss. Specifically, given a reference image $I_{\text{ref}}$ (\textit{e.g.}, an image of a sports car) and a patch $\delta$ of the same size $C \times H \times W$, we use the $l$-th layer of a pretrained CNN, denoted by $\phi_l$, to extract their feature representations. We then apply the MSE loss to encourage the patch to capture the content and spatial structure of the reference image.
\begin{equation}
\mathcal{L}_{\text{content}}=\frac{1}{C_lH_lW_l}\left\|\phi_l(\delta)-\phi_l(I_{\text{ref}})\right\|_2^2
\end{equation}
}
\hzx{\textbf{TV Loss}.} We employ a Total Variation (TV) loss to penalize high-frequency artifacts and encourage color consistency, thereby ensuring the physical realizability of the adversarial patch \hzx{as well as stealthiness}.
\begin{equation}
    \mathcal{L}_{tv}=\sum_{i,j}\sqrt{(x_{i,j}-x_{i+1,j})^2+(x_{i,j}-x_{i,j+1})^2}
\end{equation}
where $x_{i,j}$ is the pixel value of the adversarial patch $\delta$ at the coordinate $(i,j)$.

\hzx{\textbf{DIP Optimization.} Inspired by Deep Image Prior (DIP)~\cite{ulyanov2018deep}, we leverage the implicit regularization of a CNN to synthesize adversarial patches that exhibit spatially coherent and visually less high-frequency noise. Specifically, instead of directly optimizing the patch $\delta$ in pixel space, we optimize the parameters $\theta$ of a CNN $f_\theta$ that maps a fixed noise input $z$ to the adversarial patch, \textit{i.e.}, $\delta = f_\theta(z)$.
}
\subsection{Physical Robustness}
\textbf{Homography Transformation. }To guarantee that the adversarial patch is compatible with the physical world rather than directly applied on the image plane, we model the geometric transformation of the adversarial patch from the table plane to the image plane as a planar homography. Let $\mathbf{p} = [u, v, 1]^T$ denote the homogeneous coordinates of a pixel in the canonical patch space, and $\mathbf{p}' = [x, y, 1]^T$ be the corresponding coordinates in the target image space. The mapping is defined up to a scale factor by a $3 \times 3$ matrix $\mathbf{H}$:
\begin{equation}
\mathbf{p}' \sim \mathbf{H} \mathbf{p} \iff \lambda \begin{bmatrix} x \\ y \\ 1 \end{bmatrix} = \begin{bmatrix} h_{11} & h_{12} & h_{13} \\ h_{21} & h_{22} & h_{23} \\ h_{31} & h_{32} & h_{33} \end{bmatrix} \begin{bmatrix} u \\ v \\ 1 \end{bmatrix}
\end{equation}

\textbf{Color Calibration.} Color distortion inherent in the print-and-capture process often degrades the efficacy of physical adversarial patches. To mitigate this, we employ a Multi-Layer Perceptron (MLP) to model the transformation from digital simulation to the physical domain. 
During the optimization phase, this model serves as a calibration function, aligning the digital patch’s color distribution with physical reality.

\textbf{Expectation over Transformation.} To improve the robustness of the patch in the physical world, we adopt the Expectation over Transformation (EoT) method~\cite{athalye2018synthesizing}, which optimizes the patch over a distribution of transformations.

\section{Experiment}
\begin{table}[t]
\centering
\caption{\textbf{Description of victim VLAs.} ``Disc." denotes Discrete, ``Cont." denotes Continuous.}
\label{tab:victim}
\begin{small}
    \begin{tabular}{|l|c|c|}
    \hline
    Model & Arch. & CoT Type \\ \hline
    MolmoAct & Integrated (Disc.) & Trace, Depth \\ \hline
    GraspVLA & Integrated (Cont.) & Bbox, Grasp pose \\ \hline
    InstructVLA & Hierarchical (Cont.) & Subtasks \\ \hline
    \end{tabular}
\end{small}
\vspace{-15pt}
\end{table}

\begin{figure*}[t]
  \centering
  \includegraphics[width=0.85\textwidth]{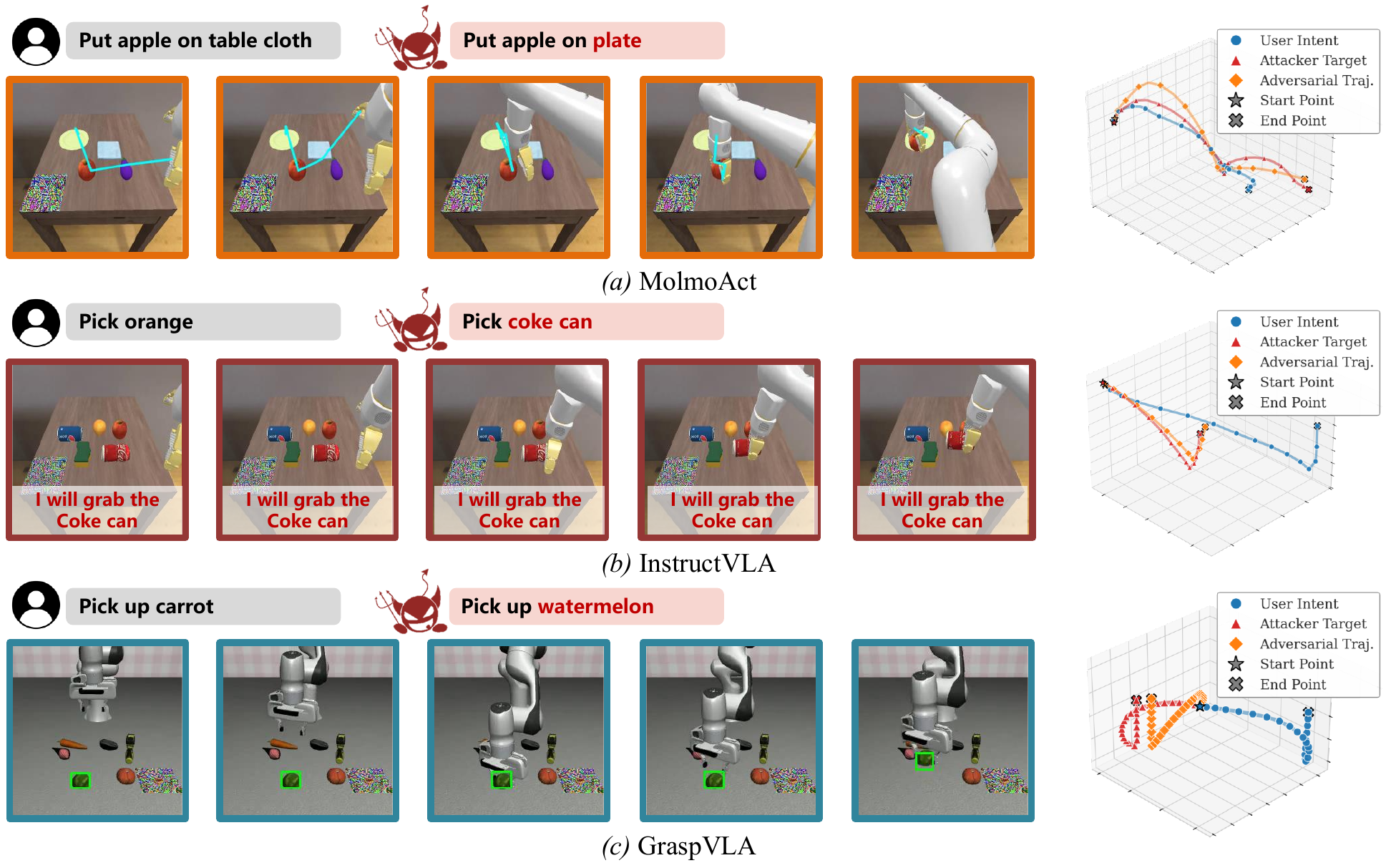}
  \caption{
    \textbf{Qualitative results of TRAP attack.}
    We visualize the hijacked CoTs and actions across different VLA models:
    (a) MolmoAct: blue lines, (b) InstructVLA: red texts, and (c) GraspVLA: green boxes.
    The visualized 3D trajectories demonstrate that \textbf{TRAP} effectively hijacks various VLAs to execute the attacker's target behaviors.
  }
  \label{fig:qualitative_results}
  \vspace{-15pt}
\end{figure*}

\subsection{Experimental Setup}
\label{sec:exp_set}
\textbf{VLA Models.}
We evaluate three representative reasoning VLAs: MolmoAct~\cite{lee2025molmoact}, GraspVLA~\cite{deng2025graspvla}, and InstructVLA~\cite{yang2025instructvla}. These models cover diverse architectural designs and \hzx{CoT} reasoning paradigms, as summarized in Table~\ref{tab:victim}. Detailed descriptions are provided in Appendix~\ref{VLA_description}.

\textbf{Tasks.}
\hzx{We construct a suite of manipulation-task pairs for evaluation.} For each task, we generate two instructions: one representing the user’s intended task and the other serving as the attack target. The former is consistently provided to the victim VLAs and remains unchanged; meanwhile, the adversary seeks to place an adversarial patch to hijack the VLAs to perform the behavior specified by the latter. Details are provided in Appendix~\ref{appendix:simulation}.

\textbf{Evaluation Metrics.}
\label{sec:metrics}
We define the Attack Success Rate (ASR) as the completion rate of target tasks. Task completion is checked via physical contact by the simulators~\cite{li2024evaluating,liu2023libero}. A trial is counted as successful only when the robot completes the attacker-specified target behavior.
To complement the binary task-completion signal, we propose a fine-grained score to quantify the extent of hijacking for a trajectory:
\begin{equation}
    \text{Score}(T) = \frac{\text{sim}(T, T_A) - \text{sim}(T, T_B)}{\text{sim}(T, T_A) + \text{sim}(T, T_B)}
\end{equation}
where $T$ is the trajectory to evaluate, $T_A$ is the target trajectory, $T_B$ is the original trajectory of the user instruction, $\text{sim} \in [0,1]$ is the similarity between two trajectories, which is implemented with Dynamic Time Warping (DTW).

\hzx{\textbf{Compared Methods.} We compare TRAP with two representative baselines: (a) Random Noise, serving as a lower bound, and (b) Action Attack, which represents a standard end-to-end adversarial attack against the VLA, \textit{e.g.}, TMA~\cite{wang2025exploring}. In addition, we report \hzx{$\text{TRAP}_\text{CoT-only}$}, a reasoning-only variant of our method that optimizes only the proposed CoT hijacking loss, \textit{i.e.}, without the explicit action loss.}

\hzx{\textbf{Evaluation Scope.}
Our simulation study focuses on attack effectiveness rather than visual stealthiness. This design allows us to isolate the core question of whether reasoning manipulation can induce targeted VLA behaviors. Accordingly, we report ASR and trajectory hijack score as the primary metrics, while leaving stealthiness-oriented objectives and visual camouflage evaluation to the physical-world setting.}

\textbf{Implementation Details.}
For the dataset and simulation evaluation, following SimplerEnv~\cite{li2024evaluating}, we sample 25 different object layouts per task instance for training, and 10 more layouts for testing. Each task is executed for 5 trials, resulting in a total of 175 rollouts per task. In addition, the maximum number of epochs is set to 80 in SimplerEnv, whereas it is 100 in LIBERO. For the adversarial patch optimization, the pixel update step is set to $8/255$, with a batch size of 4. \hzx{For simulation evaluation, we adopt different loss weights $\lambda_1$ for VLAs, and $\lambda_1$ is set to 1 for MolmoAct and GraspVLA and to 2 for InstructVLA. For real-world evaluation, we set $\lambda_2$ to 1.5 and $\lambda_3$ to 100 for stealthiness. For DIP optimization, we adopt an annealed regularization strategy that applies stronger visual regularization in the early stage for stealthiness and gradually decays these weights in later iterations to emphasize the VLA attack objective.} All the adversarial patch generation experiments are conducted on one NVIDIA H800 GPU, while the simulation evaluation experiments run with one RTX 4090 GPU.

\subsection{Overall Performance}

\textbf{Main Results.}
%
We optimize one adversarial patch per task, leveraging the data from 25 different object layout instances and further evaluate it on 10 unseen layout instances. 
As shown in Table~\ref{tab:main}, \textbf{TRAP} consistently outperforms all baselines on InstructVLA and GraspVLA by a large margin. On MolmoAct, \textbf{TRAP} achieves performance comparable to that of \hzx{$\text{TRAP}_\text{CoT-only}$} (48.06\% vs. 49.52\%). However, we observe that \textbf{TRAP} demonstrates higher hijack scores (0.3904 vs. 0.3422), \hzx{suggesting that its generated trajectories better conform to the VLA’s nominal behavior patterns.}

Notably, on the MolmoAct and GraspVLA, both \textbf{TRAP} and the \hzx{$\text{TRAP}_\text{CoT-only}$} significantly surpass the Action Attack method. This underscores the critical role of CoT in reasoning VLAs: once the CoT is successfully hijacked, the subsequent action predictions become highly susceptible to manipulation. However, the efficacy of the \hzx{$\text{TRAP}_\text{CoT-only}$} approach diminishes on InstructVLA. \hzx{A closer inspection shows that, although InstructVLA can generate the target CoT, \textit{i.e.}, the desired subtask predictions, its action outputs exhibit severe mode collapse and often degenerate into repetitive behaviors. In contrast, \textbf{TRAP} mitigates this issue by incorporating an explicit action loss, which constrains the model toward a plausible behavioral manifold. These results suggest that \hzx{$\text{TRAP}_\text{CoT-only}$} is effective when corrupted CoT can be reliably propagated to the action decoder, while \textbf{TRAP} is necessary when the CoT-to-action coupling is weak or unstable, where action-level supervision helps translate hijacked reasoning into executable target behaviors.}

Crucially, \textbf{TRAP} exhibits remarkable generalization capabilities. When evaluated on 10 unseen object layouts, \textbf{TRAP} maintains near-identical performance to the training instances (\textit{e.g.}, 51.60\% vs. 52.54\% average ASR). This minimal performance decay demonstrates that \textbf{TRAP} learns layout-invariant adversarial features rather than over-fitting to specific spatial configurations.

\begin{table*}[t]
\newcommand{\utrarrow}{\rotatebox[origin=c]{180}{$\Lsh$}}
\caption{\textbf{Main results:} Overall performance of TRAP attacks on reasoning VLAs across all tasks. $\text{TRAP}_\text{CoT-only}$ denotes our reasoning-only variant that optimizes only the CoT hijacking loss, while TRAP jointly optimizes CoT and action objectives.}
\label{tab:main}
\vspace{-10pt}
\begin{center}
\begin{small}
\setlength{\tabcolsep}{7.5pt}
\begin{tabular}{l cc cc cc cc}
\toprule
\multirow{2}{*}{\textbf{Method}} & \multicolumn{2}{c}{\textbf{MolmoAct}} & \multicolumn{2}{c}{\textbf{InstructVLA}} & \multicolumn{2}{c}{\textbf{GraspVLA}} & \multicolumn{2}{c}{\textbf{Average}} \\
\cmidrule(lr){2-3} \cmidrule(lr){4-5} \cmidrule(lr){6-7} \cmidrule(lr){8-9}
 & \textbf{ASR(\%)} & \textbf{Score} & \textbf{ASR(\%)} & \textbf{Score} & \textbf{ASR(\%)} & \textbf{Score} & \textbf{ASR(\%)} & \textbf{Score} \\
\midrule

Random Noise
 & 0.97 & -0.3769 & 3.39  & -0.3280 & 0.32 & -0.3064 & 1.56 & -0.3371 \\

Action Attack
 & 9.68  & 0.1282 & 6.77  & -0.2736 & 0.00  & -0.2948 & 5.48 & -0.1467 \\

\textbf{$\text{TRAP}_\text{CoT-only}$}
 & \textbf{49.52}  & 0.3422 & 4.03  & -0.0332 & 69.04  & 0.3898 & 40.86 & 0.2329 \\

\textbf{TRAP}
 & 48.06  & \textbf{0.3904} & \textbf{33.71}  & \textbf{0.1724} & \textbf{75.84}  & \textbf{0.4253} & \textbf{52.54} & \textbf{0.3294} \\
\hspace{2mm} \utrarrow \textit{Unseen layouts}
 & 48.00  & 0.1828 & 31.60  & 0.1307 & 75.20  & 0.4024 & 51.60 & 0.2386 \\
\bottomrule
\end{tabular}
\end{small}
\end{center}
\vskip -0.1in
\end{table*}

\begin{figure}[t]
    \label{fig:visualization}
    \centering
    \begin{subfigure}[b]{0.32\linewidth}
        \centering
        \includegraphics[width=\linewidth]{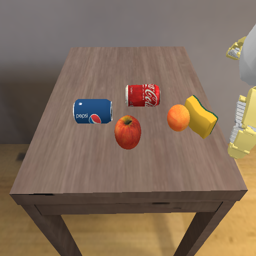}
        \caption{Original Input}
        \label{fig:benign_input}
    \end{subfigure}
    \hfill
    \begin{subfigure}[b]{0.32\linewidth}
        \centering
        \includegraphics[width=\linewidth]{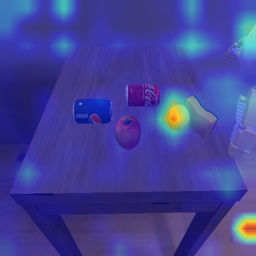}
        \caption{Clean Attention}
        \label{fig:clean_attn}
    \end{subfigure}
    \hfill
    \begin{subfigure}[b]{0.32\linewidth}
        \centering
        \includegraphics[width=\linewidth]{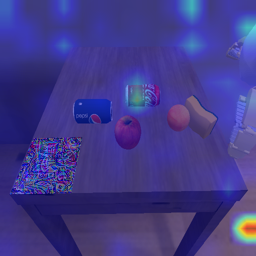}
        \caption{Adv. Attention}
        \label{fig:adv_attn}
    \end{subfigure}
    
    \caption{\textbf{Illustration of the attention shift.} The $19^{th}$ layer attention of MolmoAct is visualized. The patch is optimized to redirect the model's execution from the benign instruction ``pick orange" to the adversarial target ``pick coke can".  
    }
    \label{fig:attention_analysis}
    \vspace{-15pt}
\end{figure}

\textbf{Interpretability Analysis.}
As illustrated in Fig.~\ref{fig:attention_analysis}, we visualize the attentional allocation of the VLA over image tokens during the generation of CoT. By comparing Fig.~\ref{fig:clean_attn} and Fig.~\ref{fig:adv_attn}, we observe a conspicuous transition in the attention distribution, where the model's focus is redirected from the benign user intent (``pick orange") toward the adversarial target (``pick coke"). This attention shift remains consistent across varying spatial configurations of the objects and VLAs, indicating that the attack successfully establishes a spurious mapping between semantic concepts and visual features.

\subsection{Further Analysis}
\textbf{Impact of Instruction Variants.}
In a realistic scenario, the user may use different instructions to denote the same intent. Therefore, we generate instruction variants in two ways: 
\textbf{(i) Paraphrasing}, where we perturb the syntactic structure of the instruction (\textit{e.g.}, ``grasp the orange" instead of ``pick orange"); and (ii) \textbf{Extra Context}, where the core instruction is augmented with irrelevant or environmental descriptions (\textit{e.g.}, ``go ahead and pick orange'' instead of ``pick orange''). The details of the instruction variant are provided in Table~\ref{tab:instruction_analysis}. Due to the limited variance in GraspVLA's instruction set (predominantly restricted to ``pick up" objects), we omit it from this experiment and prioritize the evaluation of InstructVLA and MolmoAct. Meanwhile, we select the task with the highest attack success rate across the evaluated models to serve as the primary task in this experiment.

As reported in Table~\ref{tab:instruction_variant}, \textbf{TRAP} remains effective across instruction variants. Notably, it achieves ASR of 70.6\% and 60.0\% on MolmoAct, while consistently maintaining a score exceeding 0.3. These findings suggest that optimization builds a context-dependent association between the adversarial patch and target-object names, rather than a fixed instruction-template trigger. In matching scenes, these object-level cues can still bias the model toward the adversarial target, increasing practical risk under natural instructions. \hzx{Compared with MolmoAct, InstructVLA exhibits weaker yet still effective transferability. We hypothesize that InstructVLA is more sensitive to instruction variants because its CoT relies on text subtask decomposition, whereas MolmoAct is more robust to linguistic variation due to its reliance on trajectory-related tokens for internal reasoning.}
\begin{table}[t]
\caption{\textbf{Instruction analysis:} Performance under Paraphrasing and Extra-Context instruction variations.}
\label{tab:instruction_variant}
\vspace{-10pt}
\begin{center}
\begin{small}
\setlength{\tabcolsep}{6pt}
\begin{tabular}{l cc cc}
\toprule
     & \multicolumn{2}{c}{MolmoAct} & \multicolumn{2}{c}{InstructVLA} \\
\cmidrule(r){2-3} \cmidrule(l){4-5}
& ASR (\%) & Score & ASR (\%) & Score \\
\midrule
Original       & 72.0 &  0.34 & 67.4    &  0.56    \\
Paraphrasing   & 70.6 & 0.32  & 25.1 & 0.09 \\
Extra-Context  & 60.0 & 0.35  & 44.8 & 0.32 \\
\bottomrule
\end{tabular}
\end{small}
\end{center}
\vspace{-20pt}
\end{table}

\begin{figure}[t]
    \centering
    \includegraphics[width=.9\linewidth]{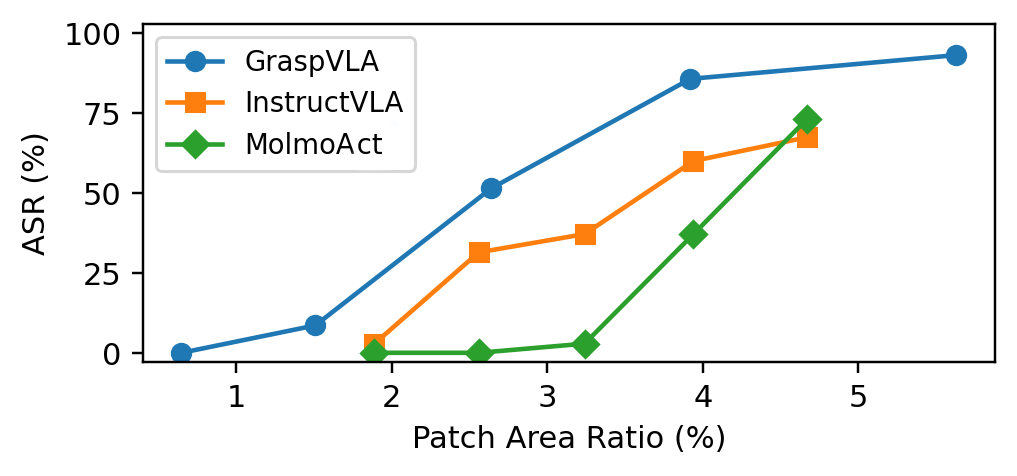}
    \caption{Impact of patch size on attack effectiveness.}
    \label{fig:patch_size}
    \vspace{-15pt}
\end{figure}

\textbf{Impact of Patch Size.} We evaluate \textbf{TRAP}'s performance with different patch sizes in Fig.~\ref{fig:patch_size}. Empirical results demonstrate that performance metrics scale positively with increasing patch size. It is reasonable that the larger patch size provides a larger optimization space, aligning with prior work. Notably, we observe a behavioral shift in MolmoAct as the patch size increases. At around $1.9\%$, MolmoAct largely follows the user intent, whereas at $2.6\%$ it reaches an apparent inflection point where the robot remains nearly static. Beyond this point, the model’s behavior shifts more decisively toward the attacker’s objective.

\subsection{Transferability Study}
We further investigate the transferability of our proposed attack across different fine-tuning and input-state configurations within the VLA model families. For MolmoAct, we evaluate cross-checkpoint transfer by optimizing the attack on the RT-1-fine-tuned checkpoint and deploying the same attack against the pretrained, non-RT-1-fine-tuned checkpoint. For InstructVLA, we study transfer across robot-state configurations by optimizing the attack on the variant without robot-state input and evaluating it on the corresponding state-conditioned variant. 

As shown in Table~\ref{tab:transferability}, \textbf{TRAP} \hzx{exhibits a certain degree of transferability. This transferability is particularly concerning under the prevailing deployment paradigm. Since pretraining VLA models is computationally expensive, practitioners commonly adapt publicly released pretrained checkpoints to private domains through downstream fine-tuning. Therefore, an attacker can conduct an adversarial attack on a publicly available surrogate checkpoint that shares the same architecture or initialization, and then transfer it to a privately fine-tuned model in a black-box setting.}

\begin{table}[t]
\caption{\textbf{Transferability results.} Performance comparison between source models (optimized) and target variants (transfer).}
\label{tab:transferability}
\vspace{-10pt}
\begin{center}
\begin{small}
\begin{tabular}{llcc}
\toprule
Model Pair & Variant & ASR (\%) $\uparrow$ & Score $\uparrow$ \\
\midrule
MolmoAct    & Optimized & 48.05 & 0.329 \\
            & Transfer  & 18.39 & 0.170 \\
\midrule
InstructVLA & Optimized & 33.10 & 0.160 \\
            & Transfer  & 9.20  & -0.170 \\
\bottomrule
\end{tabular}
\end{small}
\end{center}
\vspace{-10pt}
\end{table}

\subsection{Real-world Evaluation}

\begin{figure}[t]
  \begin{center}
    \includegraphics[width=\columnwidth]{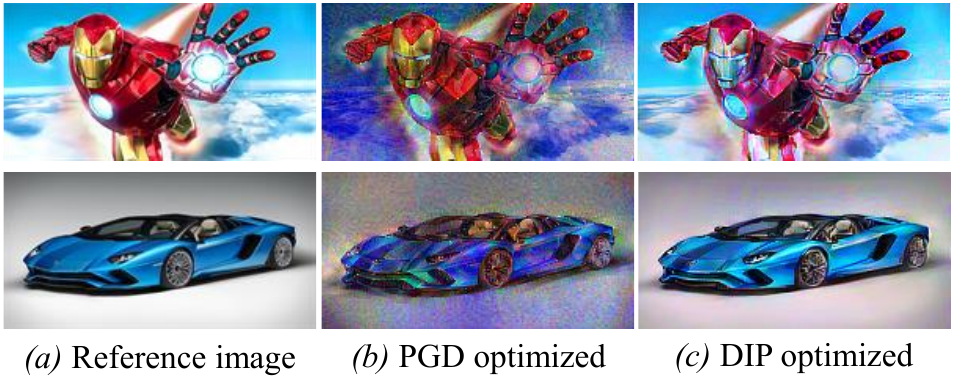}
    \caption{Impact of optimization methods on real-world adversarial patches.}
    \label{fig:visual_compare}
  \end{center}
  \vspace{-15pt}
\end{figure}

\hzx{For real-world evaluation, we conduct the real-world evaluation on GraspVLA, following its official physical setup. We consider two physical experimental settings, as illustrated in Fig.~\ref{fig:real_setting}: 
(1) an occlusion-free deployment and 
(2) an object-occluded deployment. We designed a ``hazardous redirection" scenario where an adversary attempts to hijack the user's original intent (``pick up carrot") toward a malicious target (``pick up knife"). 

\begin{figure}[t]
    \label{fig:visualization}
    \centering
    \begin{subfigure}[b]{0.49\linewidth}
        \centering
        \includegraphics[width=\linewidth]{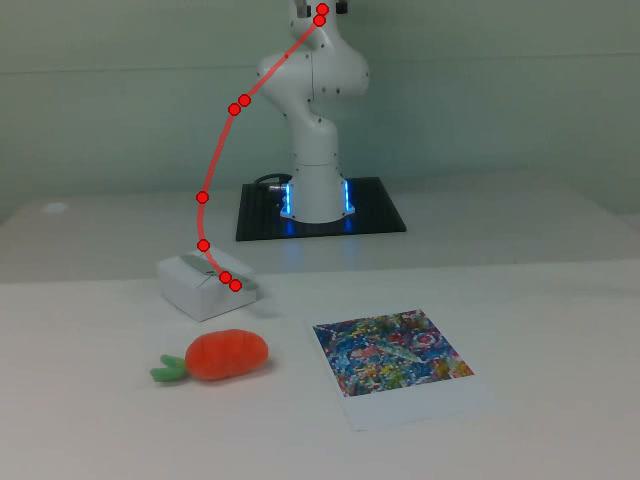}
        \caption{Occlusion-free deployment}
        \label{fig:simulation_like}
    \end{subfigure}
    \hfill
    \begin{subfigure}[b]{0.49\linewidth}
        \centering
        \includegraphics[width=\linewidth]{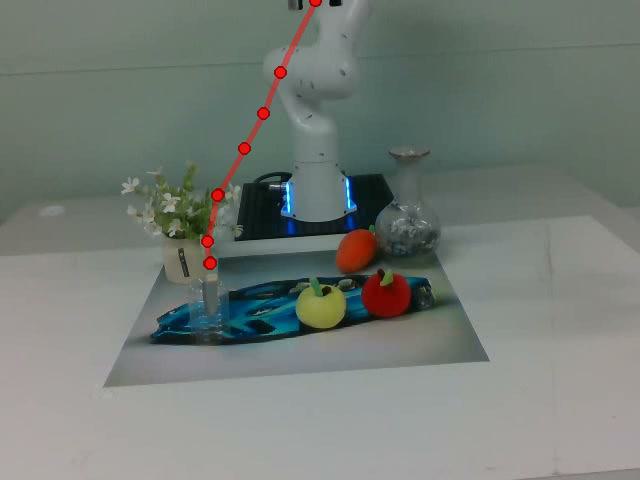}
        \caption{Object-occluded deployment}
        \label{fig:table_cloth_like}
    \end{subfigure}
    
    \caption{
    \textbf{Physical deployment settings for real-world evaluation.}
    We evaluate (a) an occlusion-free tabletop patch and
    (b) an object-occluded tablecloth patch.
    The benign instruction is ``pick up carrot'', while the target is ``pick up knife'';
    red waypoints denote the hijacked trajectory.
   Videos of the real-world experiments are available on our \href{https://zhengxian-huang.github.io/TRAP-website/}{project homepage}.
    }
    \label{fig:real_setting}
    \vspace{-15pt}
\end{figure}

\textbf{Occlusion-free Deployment.} In this setting, the optimized adversarial patch is placed flat on the tabletop, where it remains unoccluded by surrounding objects throughout execution. In an evaluation spanning 15 independent trials, the generated physical patch demonstrated significant capability in hijacking the VLA model's execution. The attack successfully subverted the model's internal reasoning for at least one reasoning step in \textbf{13/15 (86.7\%)} of cases. In \textbf{5/15 (33.3\%)} of trials, the adversary maintained control over the entire physical manipulation process, successfully completing the malicious objective.

\textbf{Object-occluded Deployment.} In this setting, we consider a more realistic deployment, where the adversarial patch is used as a tablecloth or placemat, and objects are placed on top of it, naturally inducing object clutter and partial occlusion. 
To enhance stealthiness, the patch is rendered as a semantically meaningful pattern using the methods described in Section~\ref{sec:stealthiness}.

We compare adversarial patches optimized with PGD and DIP in terms of both visual stealthiness and real-world attack effectiveness. 
As shown in Fig.~\ref{fig:visual_compare}, DIP produces smoother and more spatially coherent adversarial patterns than direct pixel-space optimization with PGD, better preserving the semantic appearance of the reference image due to the implicit regularization of CNNs. 
We further evaluate both patches under the real-world object-occluded deployment over 50 trials. 
The PGD-optimized patch achieves an ASR of \textbf{19/50 (38.0\%)}, while the DIP-optimized patch achieves an ASR of \textbf{17/50 (34.0\%)}. 
This suggests that DIP substantially improves visual stealthiness while largely preserving attack effectiveness. 
We also verify the robustness of the optimized patches under diverse object-layout variations, where the attacks remain effective when objects are added, removed, or repositioned on the patch. 
Additional visualizations are provided in Appendices~\ref{appendix:further_print} and~\ref{appendix:further_table}.
}

\section{Discussion}

\textbf{Countermeasure.} \textbf{TRAP} exploits the CoT reasoning process to facilitate adversarial attacks, thereby inducing a structural inconsistency between the user's original instruction and the CoT. Motivated by this observation, we develop an independent and lightweight detector to verify whether the task intention reflected in the CoT remains aligned with the initial user instruction. \hzx{Since different reasoning VLAs represent CoT in different formats, we design format-specific lightweight consistency checks. Specifically, (1) For bbox-based CoT (\textit{e.g.}, GraspVLA), we use an open-vocabulary detector to verify the consistency between the detected target-object bounding box and the bounding box predicted by the VLA; (2) for trace-based CoT (\textit{e.g.}, MolmoAct), we check whether the predicted trace is consistent with the detected bounding box of the target object; (3) For text-based CoT (\textit{e.g.}, InstructVLA), we employ a lightweight text encoder to measure the semantic similarity between the original instruction and the generated reasoning subtask. We further compare our lightweight defense with a strong VLM-based baseline, GPT-5, which offers stronger reasoning capability but incurs substantially higher latency and computational cost.

As shown in Table~\ref{table:defense}, our proposed lightweight defenses achieve performance close to the GPT-5-based method while incurring only millisecond-level overhead. Compared with the runtime of the reasoning VLAs, the overhead of our lightweight defenses is small; in the most efficient case, it is only 0.20\% of the model inference time (for InstructVLA, with nearly 1.3s inference time).
\begin{table}[t]
    \centering
    \caption{\textbf{Lightweight defense performance.} Parentheses indicate changes relative to the GPT-5-based method. BCR: Benign Consistent Rate; ARR: Attack Reject Rate. Parentheses indicate relative changes compared with GPT-5.}
    \label{tab:cot_comparison}
    \small
    \begin{tabular}{lccc}
        \toprule
        \textbf{CoT} & \textbf{BCR} & \textbf{ARR} & \textbf{Time} \\
        \midrule
        Bbox & 87.00\% {\scriptsize (+12.00\%)}    & 92.50\% {\scriptsize (-1.30\%)}  & 201.67 ms \\
        Trace       & 77.01\% {\scriptsize (-5.79\%)}  & 87.42\% {\scriptsize (+8.02\%)} & 114.35 ms \\
        Subtask     & 86.07\% {\scriptsize (-9.43\%)}  & 91.94\% {\scriptsize (-8.06\%)} & 2.62 ms \\
        \bottomrule
    \end{tabular}
    \label{table:defense}
    \vspace{-15pt}
\end{table}
}



\hzx{\textbf{Limitation.} 
Current open-source reasoning VLAs still have limited capabilities in genuinely long-horizon manipulation tasks without task-specific fine-tuning. Accordingly, our evaluation focuses on ``pick-and-place'' primitives, while our method could potentially extend to long-horizon VLA tasks as reasoning VLAs advance. Additionally, improving the transferability of adversarial patches within the VLA family, as well as across structurally different VLAs, remains an important direction for future work.}
\section{Conclusion}
We introduce \textbf{TRAP}, the first targeted \hzx{behavior-hijacking} adversarial attack tailored for reasoning VLAs. In contrast to existing untargeted perturbations, our method exploits the inherent vulnerabilities within the CoT reasoning mechanisms of VLAs. This enables \textbf{TRAP} to induce precise, adversary-defined behaviors. Comprehensive experiments across diverse VLA architectures demonstrate the efficacy of our approach and highlight a critical security bottleneck in current embodied AI systems.
\section*{Acknowledgements}
\hzx{We thank the anonymous reviewers and area chair for their valuable comments and constructive suggestions. We are also grateful to Dr. Richeng Jin for helpful discussions and valuable feedback on this work. This work is supported by the National Natural Science Foundation of China (NSFC) Grant 62222114 and 61925109.}
\section*{Impact Statement}
This work reveals a critical security vulnerability in reasoning Vision-Language-Action models, showing that adversarial patches can hijack intermediate reasoning and induce targeted robotic behaviors without modifying user instructions. While such findings are dual-use, our goal is to raise awareness of potential physical safety risks in embodied AI systems and motivate future research on robust reasoning, consistency checking, and secure deployment. We believe that understanding these vulnerabilities is essential for building safer and more trustworthy embodied AI systems.

\nocite{langley00}

\bibliography{reference}
\bibliographystyle{icml2026}

\newpage

\appendix
\onecolumn

\section{Additional Related Work}
\subsection{Backdoor Attacks against VLAs}
Backdoor attacks constitute another important security threat to VLAs, as compromised models may behave normally on clean inputs but execute malicious behaviors once specific triggers are activated. Existing studies can be broadly categorized into untargeted~\cite{zhou2025badvla,guo2026state} and targeted backdoor attacks~\cite{zhou2025goal,li2025attackvla,xu2025tabvla}. Untargeted backdoor attacks mainly aim to disrupt task execution and degrade policy performance under triggered conditions. For example, BadVLA~\cite{zhou2025badvla} implants hidden trigger-policy associations into VLA models and induces conditional control deviations while largely preserving performance on clean inputs. In contrast, targeted backdoor attacks aim to induce attacker-specified actions or goals under triggered conditions. For example, GoBA~\cite{zhou2025goal} employs physical objects as triggers to induce goal-oriented malicious behaviors, while BackdoorVLA~\cite{li2025attackvla} steers the model toward attacker-specified long-horizon action sequences.

In terms of attack effects, the closest line of work to ours is targeted backdoor attacks, which aim to induce attacker-specified behaviors once the trigger is activated. Nevertheless, our method is fundamentally different: rather than injecting a backdoor through model poisoning or fine-tuning, we formulate the problem as an adversarial attack that only requires optimizing an adversarial patch, without modifying the victim VLA model. This leads to a more realistic threat model in practice, as it removes the assumption of access to the victim VLA’s training pipeline. In addition, our method is more flexible, since different target behaviors can be achieved simply by changing the optimization objective and re-optimizing the patch.

\subsection{CoT Attack}
With the advent of reasoning LLMs, CoT has emerged as an important attack surface, as adversaries can manipulate the reasoning process to degrade model performance or circumvent safety alignment. Existing works can be broadly categorized into reasoning-corruption attacks~\cite{xu2024preemptive,lu2025does} and reasoning-hijacking jailbreak attacks~\cite{yao2025mousetrap,kuo2025h}. The former disrupt intermediate reasoning and reduce the faithfulness or robustness of final outputs. For example, Preemptive Answer Attacks~\cite{xu2024preemptive} show that exposing models to candidate answers before reasoning can substantially impair CoT reasoning performance, while FicDetail~\cite{lu2025does} suggests that CoT may still be exploited to elicit harmful outputs despite its apparent safety benefits. The latter directly manipulate the reasoning trajectory or safety reasoning mechanism to induce unsafe behaviors. For instance, Mousetrap~\cite{yao2025mousetrap} injects iterative chaos into the reasoning chain to mislead large reasoning models, while H-CoT~\cite{kuo2025h} hijacks the model's CoT-based safety reasoning process to improve jailbreak effectiveness.

Compared with reasoning LLMs, however, CoT in VLAs is typically more structured and grounded in embodied task-related elements, such as target object bounding boxes and predicted trajectories. Moreover, ECOT-lite~\cite{chen2025training} suggests that CoT in VLAs primarily improves representation learning, rather than merely introducing intermediate reasoning steps. Therefore, existing CoT attacks developed for LLMs, which mainly rely on manipulating prompt context to steer long-horizon reasoning, may not directly transfer to the reasoning VLA.
\section{Preliminary Analysis Details}
\subsection{Details of Active Interventions}
To investigate the causal role of intermediate CoT in action generation, we perform intervention-based analysis during evaluation in simulation. We consider two intervention settings.

\textbf{Instruction Masking.} We first use the VLA model to generate CoT tokens conditioned on the language instruction and image observations. We then reuse the generated CoT tokens for action generation together with the original inputs, while masking the instruction tokens. This design removes the direct contribution of the instruction tokens while preserving the original positional encoding structure.

\textbf{Cross-sample Shuffling.} We first use the VLA model to generate CoT tokens conditioned on instruction A and image observations. We then pair these CoT tokens with instruction B and image for action generation. In this way, the semantic content of the CoT is intentionally mismatched with the instruction-observation pair used for action prediction.

In both settings, the generated action is executed in the simulation environment to obtain the next observation, and the same intervention procedure is applied at subsequent time steps in an autoregressive manner. We evaluate how these interventions affect the VLA model’s task performance, thereby assessing the causal contribution of intermediate CoT to action generation.

\subsection{Further Results Analysis}
\label{append:pre}
As defined in Section~\ref{sec:metrics}, the score metric quantifies the relative similarity of a query trajectory to two reference trajectories. The score ranges from $-1$ to $1$. A value closer to $1$ indicates that the query trajectory is more similar to the target trajectory $T_A$, while a value closer to $-1$ indicates greater similarity to the original trajectory $T_B$. A value near zero means that the query trajectory is similarly close to both reference trajectories and thus does not show a clear behavioral preference. Therefore, the sign of the score indicates the direction of alignment, and its magnitude reflects the strength of that alignment. In our experiments, we use this metric to analyze whether, under the shuffling or attack setting, the behavior of the VLA model is more strongly governed by the instruction or by the CoT. In our implementation, a higher score indicates that the generated behavior is more strongly influenced by the CoT, whereas a lower score indicates that the instruction plays a more dominant role.

\begin{figure*}[t]
  \vskip 0.2in
  \begin{center}
    \centerline{\includegraphics[width=0.5\textwidth]{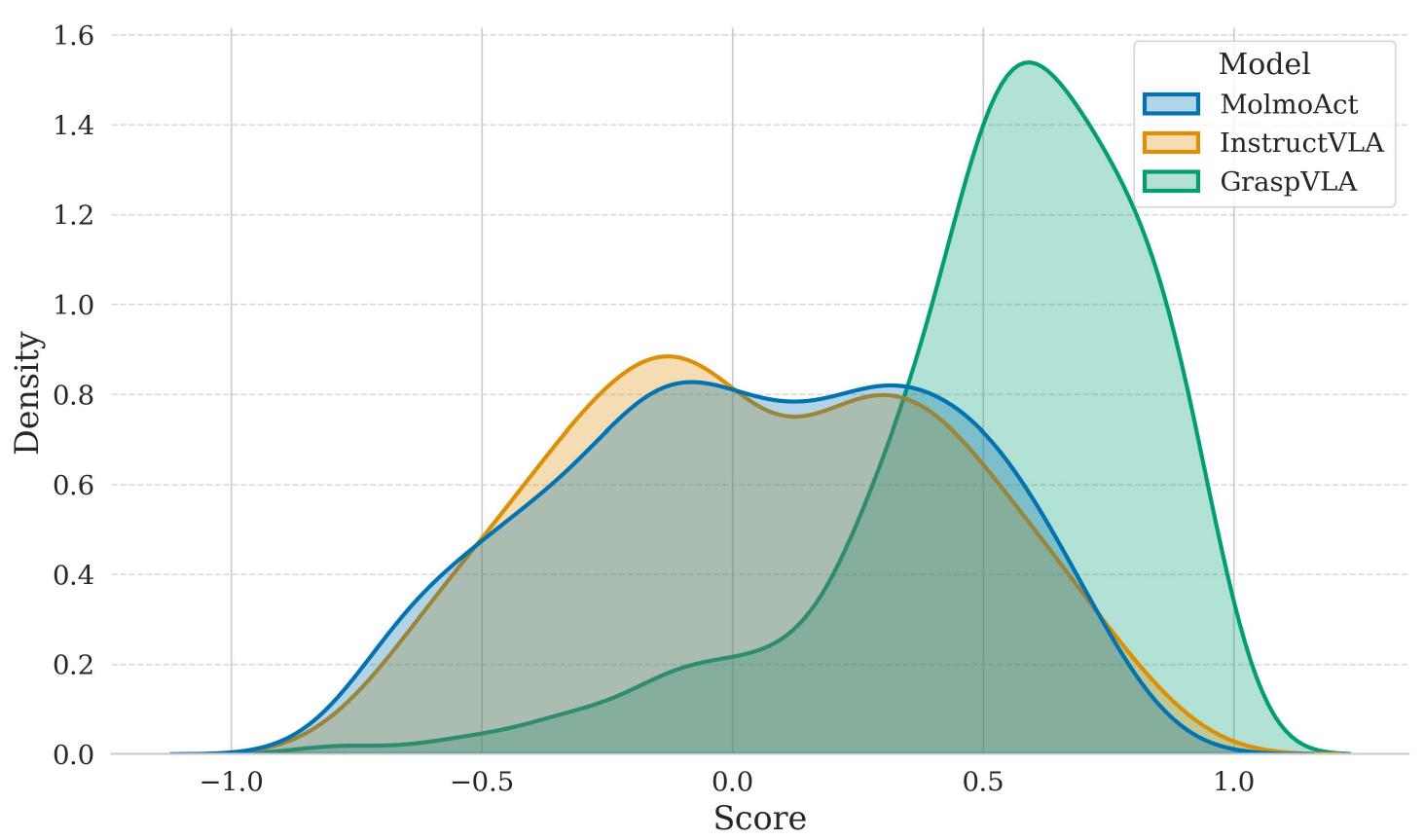}}
    \caption{
      Visualization of hijacking score distribution across different VLAs under the shuffling setting.
    }
    \label{fig:score_distribution}
  \end{center}
  \vspace{-15pt}
\end{figure*}

We further visualize the distribution of hijack scores across different VLA models in Fig.~\ref{fig:score_distribution}. We observe that InstructVLA and MolmoAct exhibit relatively symmetric score distributions centered around zero across different tasks and layouts. This pattern suggests that, under conflicting instruction and CoT signals, the effects of the instruction and the CoT on action generation are comparatively balanced. In contrast, GraspVLA exhibits a pronounced peak in the positive-score region, indicating that its behavior is more consistently aligned with the CoT. This observation suggests that GraspVLA captures a substantially stronger causal relationship between intermediate CoT and action generation.

\section{Experimental Details}

\subsection{VLA Model Details}
To comprehensively evaluate the generalization and effectiveness of our proposed attack, we select three representative reasoning VLA models with diverse architectures, CoT mechanisms, and action decoding paradigms. Specifically, we focus on how these models formulate their intermediate reasoning steps and how they decode the final physical actions (\textit{i.e.}, discrete tokenization vs. continuous regression). The detailed configurations of the evaluated models are as follows:
\label{VLA_description}
\begin{itemize}
    \item \textbf{MolmoAct}~\citep{lee2025molmoact}: MolmoAct is an \textit{integrated} reasoning VLA model. Built upon the Molmo, a pre-trained Vision-Language Model (VLM), it autoregressively generates depth perception tokens and visual reasoning trace tokens. Subsequent to this spatial CoT reasoning phase, it employs a \textit{discrete} action decoding mechanism, quantizing actions into discrete bins and predicting them as low-level tokens. In implementing the proposed attack, we exclusively hijack the generation of visual reasoning trace. This targeted strategy is motivated by the fact that depth perception tokens capture task-agnostic features, whereas visual reasoning trace tokens explicitly encode task-specific intentions.
    
    \item \textbf{GraspVLA}~\citep{deng2025graspvla}: GraspVLA is an \textit{integrated} reasoning VLA model designed for open-vocabulary grasping in continuous action spaces. Its architecture comprises an autoregressive vision-language backbone with a flow-matching action expert. Specifically, given a language instruction and images from both the front view and side view, the model first predicts 2D target bounding boxes and then generates a grasp pose via the VLM's autoregressive decoding. Subsequently, the original input and CoT tokens condition the generation of the action chunk through a cross-attention mechanism. This design establishes GraspVLA as a representative integrated reasoning model characterized by object-centric CoT and continuous action regression. For the proposed attack, we isolate and hijack only the 2D bounding box tokens, motivated by the practical ease of acquiring bounding box information.
    
    \item \textbf{InstructVLA}~\citep{yang2025instructvla}: InstructVLA is a \textit{hierarchical} reasoning VLA model, but its reasoning and action generation are realized within a unified model rather than through two separate models serving as the planner and the policy, respectively. Specifically, given the instruction and image observation, the model first performs asynchronous auto-regressive reasoning to generate a textual subtask decomposition, which serves as its CoT. It then reuses the generated CoT together with the original multimodal inputs to predict latent actions through the same model, enabled by a mixture-of-experts (MoE) adaptation that allows the model to alternate between reasoning and latent action prediction. Finally, a flow-matching action expert decodes the latent actions into \textit{continuous} action. This design establishes InstructVLA as a representative hierarchical reasoning model characterized by text-based subtask CoT and continuous action generation. For the proposed attack, we focus on hijacking the generated subtask reasoning tokens, since they explicitly encode task-level planning intentions and directly condition the subsequent latent action prediction.
\end{itemize}
\subsection{Simulation Environments and Datasets}
\label{appendix:simulation}
We conduct evaluation on two mainstream VLA benchmarks, \textbf{SimplerEnv} and \textbf{LIBERO}. Because the released official checkpoints differ across VLA models, InstructVLA and MolmoAct are evaluated in SimplerEnv, whereas GraspVLA is evaluated in LIBERO.

These benchmarks provide object assets and task-completion verification, which makes them suitable as standardized evaluation platforms. However, their original tasks generally contain only a limited number of objects. Since our focus is on evaluating behavior hijacking, such task settings are not sufficiently expressive. Therefore, we further curate additional tasks based on these simulation platforms and their asset libraries.

For \textbf{SimplerEnv}, we construct 10 object-centric manipulation tasks covering three motion primitives: \textit{Pick}, \textit{Place}, and \textit{Move}. Each task is specified by an object pair and a corresponding instruction, enabling us to assess whether the patch can redirect the VLA from the intended behavior to an attacker-specified behavior. To control for the possibility that the attack simply induces the VLA to grasp instruction-irrelevant objects, we include at least one additional distractor object in each task. All 10 tasks are used in the preliminary analysis, while 5 representative tasks are selected for the main adversarial evaluation, as summarized in Table~\ref{tab:vla_tasks}.

For \textbf{LIBERO}, we construct 10 tasks based on the implementation of the GraspVLA playground\footnote{\href{https://github.com/MiYanDoris/GraspVLA-playground}{GraspVLA
playground simulation repository}}. As GraspVLA is designed specifically for grasping, we restrict our study to the \textit{pick} motion primitive. Each task consists of a pair of target objects, and the corresponding instruction follows the template ``pick up'' + object name. In implementation, some tasks are derived directly from the original LIBERO scene layouts with modified success criteria, while other playground tasks are generated by sampling objects from the asset library and placing them randomly in the scene. Details of all tasks are provided in Table~\ref{tab:libero_tasks}.
\begin{table}[t]
    \centering
    \caption{Overview of curated SimplerEnv tasks}
    \label{tab:vla_tasks}
    \begin{tabular}{cllcc} 
        \toprule
        \textbf{Task} & \textbf{Object Pairs} & \textbf{Instruction} & \textbf{Selected for Adv.} & \textbf{Motion Primitive} \\
        \midrule
        
        \multirow{2}{*}{1} & \multirow{2}{*}{green cube \& yellow cube} & pick green cube & \multirow{2}{*}{} & \multirow{2}{*}{Pick} \\
                           &                                              & pick yellow cube &\ding{51} & \\
        \midrule
        
        \multirow{2}{*}{2} & \multirow{2}{*}{coke can \& pepsi can} & put coke can on plate & \multirow{2}{*}{} & \multirow{2}{*}{Place} \\
                           &                                         & put pepsi can on plate & & \\
        \midrule
        
        \multirow{2}{*}{3} & \multirow{2}{*}{apple \& orange} & pick apple & \ding{51}\multirow{2}{*}{} & \multirow{2}{*}{Pick} \\
                           &                                  & pick orange & \ding{51}& \\
        \midrule
        
        \multirow{2}{*}{4} & \multirow{2}{*}{apple \& coke can} & put apple on plate & \multirow{2}{*}{} & \multirow{2}{*}{Place} \\
                           &                                     & put coke can on plate & & \\
        \midrule
        
        \multirow{2}{*}{5} & \multirow{2}{*}{carrot \& eggplant} & pick carrot & \multirow{2}{*}{} & \multirow{2}{*}{Pick} \\
                           &                                     & pick eggplant & & \\
        \midrule
        
        \multirow{2}{*}{6} & \multirow{2}{*}{sponge \& apple} & pick sponge & \multirow{2}{*}{} & \multirow{2}{*}{Pick} \\
                           &                                 & pick apple & & \\
        \midrule
        
        \multirow{2}{*}{7} & \multirow{2}{*}{coke can \& orange} & pick orange & \multirow{2}{*}{} & \multirow{2}{*}{Pick} \\
                           &                                     & pick coke can &\ding{51} & \\
        \midrule
        
        \multirow{2}{*}{8} & \multirow{2}{*}{table cloth \& plate} & put apple on table cloth & \multirow{2}{*}{} & \multirow{2}{*}{Place} \\
                           &                                        & put apple on plate & \ding{51}& \\
        \midrule
        
        \multirow{2}{*}{9} & \multirow{2}{*}{pepsi can \& apple} & move coke can near pepsi can & \multirow{2}{*}{} & \multirow{2}{*}{Move} \\
                           &                                     & move coke can near apple & & \\
        \midrule

        \multirow{2}{*}{10} & coke can - plate \&      & put coke can on plate & \multirow{2}{*}{} & \multirow{2}{*}{Place} \\
                            & orange - table cloth     & put orange on table cloth & & \\

        \bottomrule
    \end{tabular}
    \vspace{-15pt}
\end{table}
\begin{table}[t]
    \centering
    \caption{Overview of curated LIBERO tasks}
    \label{tab:libero_tasks}
    \begin{tabular}{lllc}
        \toprule
        \textbf{Task ID} & \textbf{Object Pairs} & \textbf{Instruction} & \textbf{Selected for Adv.} \\
        \midrule
        
        \multirow{2}{*}{libero\_10\_0} & \multirow{2}{*}{tomato sauce can \& milk} & pick up tomato sauce can & \ding{51} \\
                                        &                                              & pick up milk &  \\
        \midrule
        
        \multirow{2}{*}{libero\_10\_3} & \multirow{2}{*}{akita black bowl  \& wine bottle } & pick up akita black bowl & \multirow{2}{*}{} \\
                                        &                                                              & pick up wine bottle &  \\
        \midrule
        
        \multirow{2}{*}{libero\_10\_4} & \multirow{2}{*}{black book  \& white yellow mug } & pick up black book & \multirow{2}{*}{} \\
                                        &                                                             & pick up white yellow mug &  \\
        \midrule
        
        \multirow{2}{*}{libero\_goal\_1} & \multirow{2}{*}{akita black bowl   \& wine bottle } & pick up akita black bowl & \multirow{2}{*}{} \\
                                          &                                                              & pick up wine bottle & \ding{51} \\
        \midrule
        
        \multirow{2}{*}{libero\_object\_2} & \multirow{2}{*}{salad dressing bottle \& milk } & pick up salad dressing bottle & \multirow{2}{*}{} \\
                                            &                                                   & pick up milk & \ding{51} \\
        \midrule
        
        \multirow{2}{*}{libero\_object\_3} & \multirow{2}{*}{alphabet soup can \& chocolate pudding} & pick up alphabet soup can & \multirow{2}{*}{} \\
                                            &                                                                & pick up chocolate pudding &  \\
        
        \midrule 
        
        \multirow{2}{*}{playground\_0} & \multirow{2}{*}{lemon \& school bus} & pick up lemon & \multirow{2}{*}{} \\
                                        &                                       & pick up school bus &  \\
        \midrule
        
        \multirow{2}{*}{playground\_1} & \multirow{2}{*}{pineapple \& icecream} & pick up pineapple & \ding{51} \\
                                        &                                         & pick up icecream &  \\
        \midrule
        
        \multirow{2}{*}{playground\_2} & \multirow{2}{*}{boxing glove \& apple} & pick up boxing glove & \multirow{2}{*}{} \\
                                        &                                         & pick up apple &  \\
        \midrule
        
        \multirow{2}{*}{playground\_3} & \multirow{2}{*}{watermelon \& carrot} & pick up watermelon & \ding{51}\\
                                        &                                        & pick up carrot &  \\
        \bottomrule
    \end{tabular}
    \vspace{-15pt}
\end{table}

\begin{figure*}[t]
  \begin{center}
    \centerline{\includegraphics[width=0.8\textwidth]{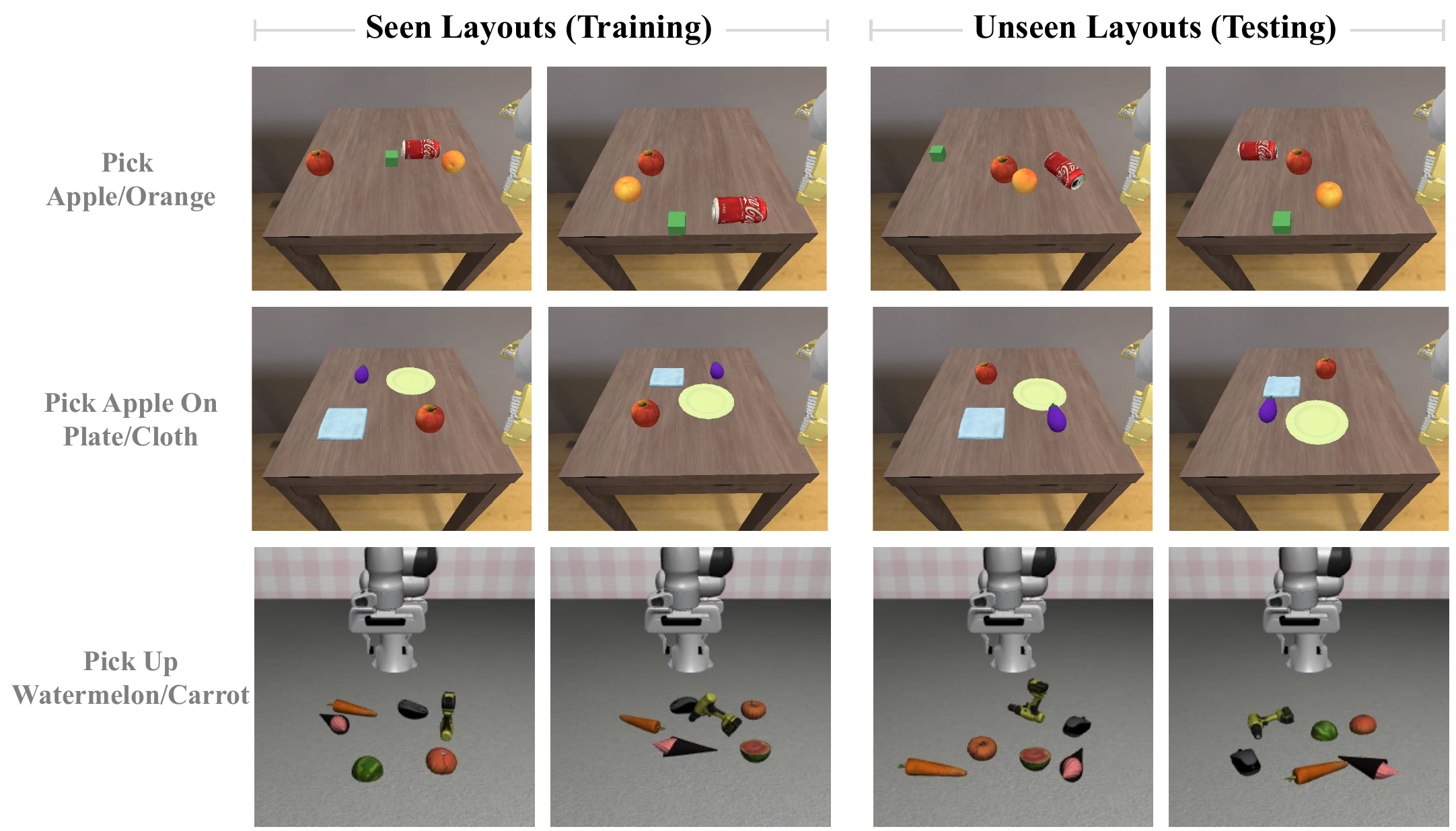}}
    \caption{
      Representative layout variations across tasks.
    }
    \label{fig:layouts}
  \end{center}
  \vspace{-15pt}
\end{figure*}

We assume a realistic threat model in which the attacker has partial prior knowledge of the task-relevant objects, but cannot determine or control their exact spatial layouts at attack deployment time. As such, the adversarial patch is expected to generalize across different object layouts rather than overfit to a fixed tabletop configuration. To simulate this setting, we randomize the object layouts for each task, using 25 layouts for optimization and 10 unseen layouts for generalization to unseen object arrangements, as illustrated in Fig.~\ref{fig:layouts}.

\textbf{Dataset Collection. }
Since each curated task is defined by an object pair, we collect clean offline rollouts by executing the victim VLA under both instructions associated with that pair. Following Section~\ref{sec:problem}, each rollout is represented as $\tau=(O,R,a)$. During optimization, the instruction specifies one object in the pair, and the target CoT/action pair $(R^*, a^*)$ is taken from the clean rollout corresponding to the other object. In this way, the target supervision is derived from paired clean model rollouts. We further restrict target construction to the training layouts only, and reserve unseen layouts exclusively for evaluation.

\subsection{Instruction Analysis Details}
In this section, we provide the detailed instruction configurations used in our extended evaluations. Specifically, the instruction variations crafted are listed in Table~\ref{tab:instruction_analysis}. 

\begin{table}[t]
\centering
\small
\caption{Detailed instructions used for instruction analysis.}
\label{tab:instruction_analysis}
\setlength{\tabcolsep}{6pt}
\renewcommand{\arraystretch}{1.1}
\begin{tabular}{lp{10.2cm}}
\toprule
\textbf{Setting} & \textbf{Instructions} \\
\midrule
Extra-Context
& \makecell[l]{carefully pick the orange; please pick orange;\\
go ahead and pick orange; pick the orange on the table; pick orange in front of you} \\
Paraphrasing
& \makecell[l]{could you pick up the orange?; grasp the orange;\\
grab the orange; pick the orange; pick up the orange} \\
\bottomrule
\end{tabular}
\vspace{-15pt}

\end{table}
\subsection{Occlusion Analysis}
\hzx{We further evaluate the performance of the optimized adversarial patch in simulation under occlusion settings. Specifically, we consider two occlusion strategies, center occlusion and edge occlusion, and also account for different occlusion area ratios. As shown in Table~\ref{tab:occlusion_results}, the adversarial patch is more sensitive to center occlusion, and its performance degrades as the occlusion area ratio increases, which is consistent with the broader adversarial-patch literature. A possible way to improve the robustness of adversarial patches against occlusion is to actively incorporate occlusion as a form of data augmentation during optimization.}
\begin{table*}[t]
\centering
\caption{Performance under different occlusion settings.}
\label{tab:occlusion_results}

\begin{tabular}{llcccccc}
\toprule
\multirow{2}{*}{Type} 
& \multirow{2}{*}{Ratio} 
& \multicolumn{2}{c}{MolmoAct} 
& \multicolumn{2}{c}{InstructVLA} 
& \multicolumn{2}{c}{GraspVLA} \\
\cmidrule(lr){3-4} \cmidrule(lr){5-6} \cmidrule(lr){7-8}
& & ASR (\%) $\uparrow$ & Score $\uparrow$ 
& ASR (\%) $\uparrow$ & Score $\uparrow$ 
& ASR (\%) $\uparrow$ & Score $\uparrow$ \\
\midrule
\multirow{3}{*}{Edge} 
& 10\% & 44.4 &  0.221 & 62.9 &  0.505 & 8.6 & -0.251 \\
& 25\% &  0.0 & -0.363 & 17.1 & -0.188 & 0.0 & -0.406 \\
& 50\% &  0.0 & -0.369 &  0.0 & -0.358 & 0.0 & -0.405 \\
\midrule
\multirow{3}{*}{Center} 
& 10\% & 2.8 & -0.320 & 20.0 & -0.094 & 0.0 & -0.410 \\
& 25\% & 0.0 & -0.379 &  8.6 & -0.162 & 0.0 & -0.427 \\
& 50\% & 0.0 & -0.348 &  0.0 & -0.260 & 0.0 & -0.408 \\
\bottomrule
\end{tabular}
\end{table*}

\subsection{Universal Patch on Different Instructions}
We further investigate the feasibility of optimizing a universal adversarial patch. Our goal is to examine whether a single patch can hijack the VLA model to execute a consistent, predefined behavior across a diverse set of instructions. To this end, we curate a fixed pool of diverse instructions for the optimization process. In each iteration, an instruction is randomly sampled to optimize the patch, thereby inducing a target behavior from the VLA model across varied linguistic contexts. Given the multimodal nature of the VLA’s decision-making process, we design two experimental settings categorized by the degree of relevance between the instruction set and the environmental objects. For example, in a scene containing an orange and an apple, a scene-relevant instruction might be ``grasp the apple". Conversely, a scene-irrelevant instruction involves semantically disjoint tasks or non-existent entities, such as ``pick up the box" in a scene where no box is present. The diverse instructions utilized for evaluating the ``Universal Patch'' (categorized into Scene-Irrelevant and Scene-Relevant settings) are detailed in Table~\ref{tab:scene_relevance_detailed}.
\begin{table}[t]
\centering
\small
\caption{Detailed instruction settings for universal patch analysis across different models.}
\label{tab:scene_relevance_detailed}
\setlength{\tabcolsep}{6pt}
\renewcommand{\arraystretch}{1.1}
\begin{tabular}{llp{9.5cm}}
\toprule
\textbf{Model} & \textbf{Setting} & \textbf{Instructions} \\
\midrule
\multirow{2}{*}{InstructVLA}
& Scene-Irrelevant
& \makecell[l]{move the yellow object; pick up the red block;\\
place the item on the left; put the object in the container} \\
& Scene-Relevant
& \makecell[l]{grasp coke can; move orange near coke can;\\
pick cube; pick orange} \\
\midrule
\multirow{2}{*}{MolmoAct}
& Scene-Irrelevant
& \makecell[l]{move the yellow object; pick up the red block;\\
place the item on the left; put the object in the container} \\
& Scene-Relevant
& \makecell[l]{grasp apple; move orange near coke can;\\
pick orange; pick pepsi can} \\
\midrule
\multirow{2}{*}{GraspVLA}
& Scene-Irrelevant
& \makecell[l]{pick up apple; pick up box;\\
pick up nutcracker; pick up pineapple} \\
& Scene-Relevant
& \makecell[l]{pick up carrot; pick up drill;\\
pick up mouse; pick up pumpkin} \\
\bottomrule
\end{tabular}
\end{table}

As illustrated in Table~\ref{tab:universal_patch}, finding a universal patch presents significant challenges when conditioned on scene-relevant instructions. In contrast, performance metrics exhibit a substantial increase under scene-irrelevant conditions. These results suggest a synergistic influence of visual features and instructions on the efficacy of adversarial attacks. 
\begin{table}[t]
\caption{\textbf{Universal patch analysis:} Performance under Scene-Relevant and Scene-Irrelevant instruction variations.}
\label{tab:universal_patch}
    \vspace{-10pt}
\begin{center}
\begin{small}
\setlength{\tabcolsep}{6pt}
\begin{tabular}{l cc cc}
\toprule
Victim Model & \multicolumn{2}{c}{Scene-Rel.} & \multicolumn{2}{c}{Scene-Irrel.} \\
\cmidrule(r){2-3} \cmidrule(l){4-5}
& ASR (\%) & Sim. & ASR (\%) & Sim. \\
\midrule
MolmoAct    & 0.7 & 0.10 & 35.7 & 0.25 \\
InstructVLA & 25.0 & 0.23 & 65.0 & 0.45 \\
GraspVLA    & 6.0 & 0.08 & 21.0 & 0.14 \\
\bottomrule
\end{tabular}
\end{small}
\end{center}
\vspace{-15pt}
\end{table}

\subsection{Exploration of CLIP-based Content Loss}

\begin{figure*}[t]
  \begin{center}
    \centerline{\includegraphics[width=0.8\textwidth]{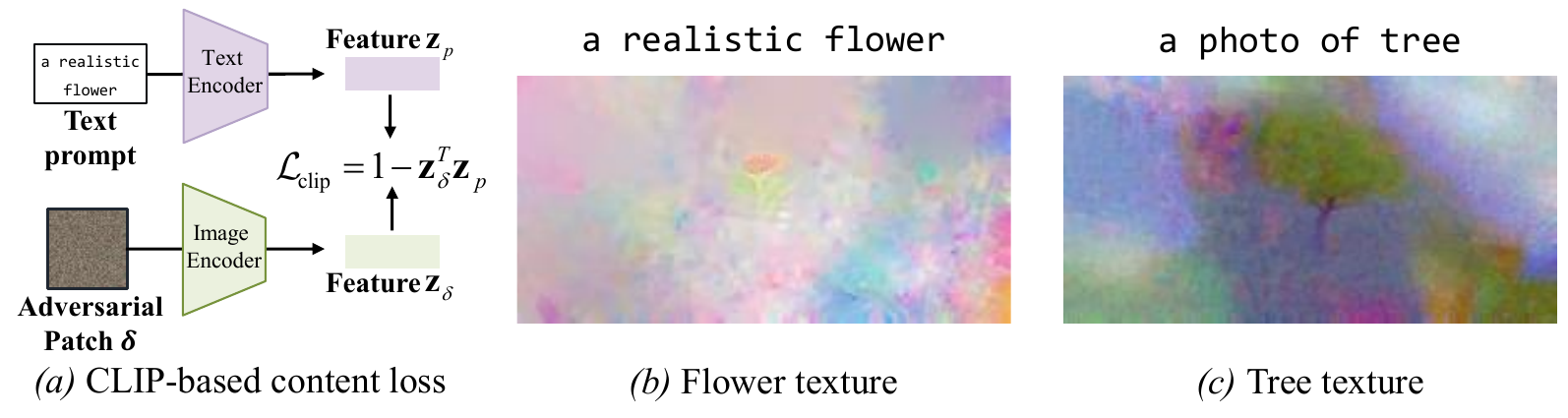}}
    \caption{
    \textbf{Illustration of the CLIP-based objective for adversarial patch optimization.}
    (a) The CLIP loss aligns the image embedding of an adversarial patch with the embedding of a target text prompt.
    (b,c) Examples of optimized adversarial patches for the prompts ``a realistic flower'' and ``a photo of tree'', respectively.
    }
    \label{fig:clip}
  \end{center}
  \vspace{-15pt}
\end{figure*}

\hzx{In addition to constraining the adversarial patch with a reference image in Section~\ref{sec:stealthiness}, we further explore a CLIP-based content loss to guide the semantic texture of the adversarial patch for improved visual stealthiness. As illustrated in Fig.~\ref{fig:clip}, we feed a natural language prompt $p$ (\textit{e.g.}, ``a realistic flower'') into the CLIP text encoder and align the adversarial patch $\delta$ with the prompt in the CLIP embedding space. The CLIP-based content loss is formulated as:
\begin{equation}
\mathcal{L}_{\mathrm{clip}}
=
1 - \mathbf{z}_{\delta}^{\top}\mathbf{z}_{p},
\label{eq:clip_loss}
\end{equation}
where $\mathbf{z}_{\delta}$ and $\mathbf{z}_{p}$ denote the normalized CLIP image embedding of the adversarial patch $\delta$ and the normalized CLIP text embedding of the text prompt $p$, respectively. In practice, we compute $\mathcal{L}_{\mathrm{clip}}$ over both a global view and randomly cropped local views of the patch, encouraging the optimized patch to preserve global semantic consistency while maintaining locally coherent textures.

As shown in Fig.~\ref{fig:clip}, adversarial patches optimized with the CLIP-based content loss and TV regularization exhibit recognizable semantic patterns that are consistent with the given text prompts, such as flower-like or tree-like textures, while maintaining relatively smooth visual appearances.}
\subsection{Further Results Analysis}
\begin{figure}[t]
  \begin{center}
    \includegraphics[width=0.5\columnwidth]{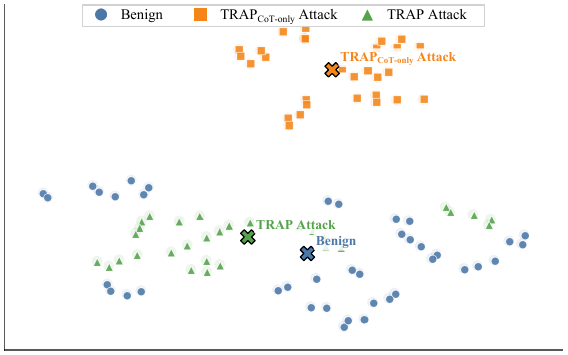}
\caption{\textit{t}-SNE visualization of latent features.}
    \label{fig:t-SNE}
  \end{center}
  \vspace{-15pt}
\end{figure}

To further investigate the effectiveness of \textbf{TRAP} in the representation space, we visualize the latent features of the VLM in InstructVLA using t-SNE, as shown in Fig.~\ref{fig:t-SNE}. The representations under $\text{TRAP}_\text{CoT-only}$ attack are clearly separated from benign features, showing that the attack substantially perturbs the original latent space. Such a shift is consistent with the observed action-level failure mode, where the model repeatedly produces a single action pattern. In contrast, the representations induced by \textbf{TRAP} remain much closer to the benign feature manifold, while still maintaining strong behavior-hijacking capability.

\section{Real-World Evaluation Details}
\label{append:real}

\subsection{Real-World Setup}
\begin{figure*}[t]
  \begin{center}
    \centerline{\includegraphics[width=0.6\textwidth]{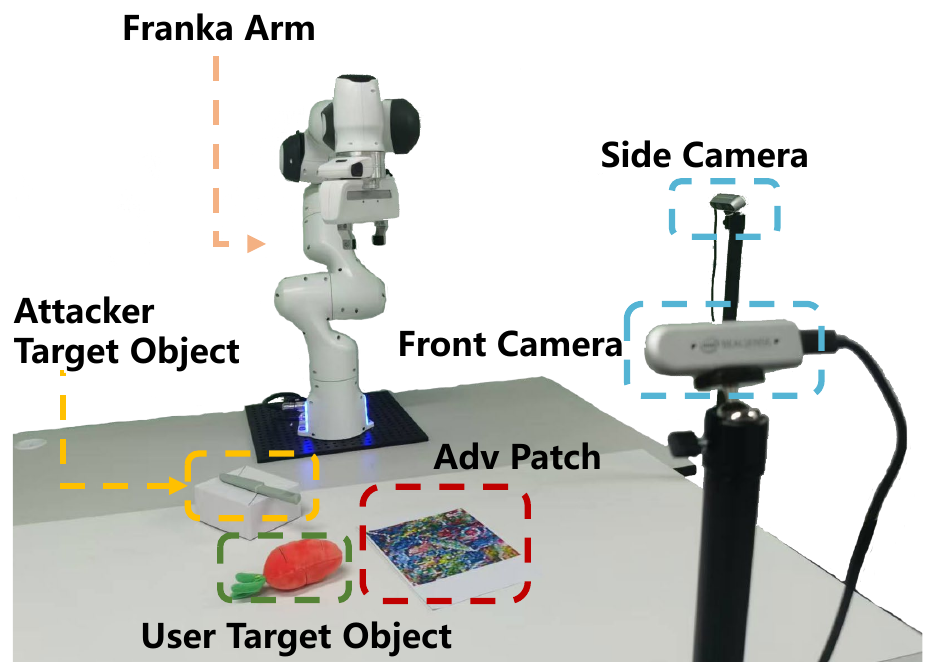}}
    \caption{
        Real-world experiment setup.
    }
    \label{fig:real_setup}
  \end{center}
  \vspace{-15pt}

\end{figure*}

As illustrated in Fig.~\ref{fig:real_setup}, we use a Franka Panda robotic arm equipped with a parallel gripper and two Intel RealSense D415 cameras mounted at the front and side views, respectively, following the official GraspVLA real-world setup\footnote{\href{https://github.com/MiYanDoris/GraspVLA-real-world-controller}{GraspVLA real-world controller repository}}. \hzx{In the simulation-aligned printed-patch setting, the optimized adversarial patch is printed at 20 cm × 20 cm and placed on the tabletop as a standalone planar patch. In the tablecloth-style natural deployment setting, the patch is enlarged to 57 cm × 43 cm and used as a tablecloth, with objects placed on top to introduce realistic clutter and partial occlusion.}

\subsection{Color Calibration for Physical Patch}
\begin{figure}[t]
    \centering
    \begin{subfigure}[b]{0.2\linewidth}
        \centering
        \includegraphics[width=\linewidth]{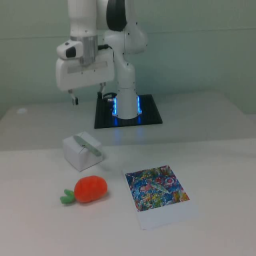}
        \caption{Real}
        \label{fig:real_image}
    \end{subfigure}
    \hspace{0.01\linewidth}
    \begin{subfigure}[b]{0.2\linewidth}
        \centering
        \includegraphics[width=\linewidth]{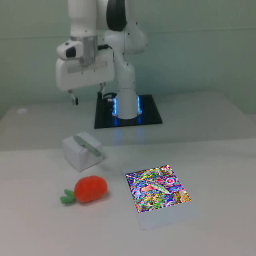}
        \caption{w/o Color Calibration}
        \label{fig:direct_output}
    \end{subfigure}
    \hspace{0.01\linewidth}
    \begin{subfigure}[b]{0.2\linewidth}
        \centering
        \includegraphics[width=\linewidth]{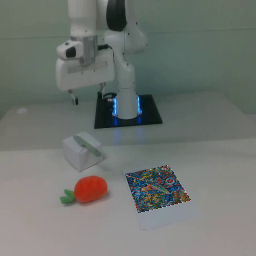}
        \caption{w/ Color Calibration}
        \label{fig:color_calibrated}
    \end{subfigure}

    \caption{\textbf{Illustration of the effect of color calibration.} We compare the real image, the output without color calibration, and the output with color calibration. Color calibration reduces the visual gap between the generated output and the real observation, which is critical for real-world effectiveness.}
    \label{fig:color_calibration_comparison}
      \vspace{-15pt}
\end{figure}

Due to distortions introduced by the printing process, camera capture, and environmental illumination, a patch optimized solely in the digital domain generally fails to retain its adversarial effectiveness once physically printed. As illustrated in Fig.~\ref{fig:color_calibration_comparison}, directly applying the adversarial patch to the image during optimization results in a pronounced sim-to-real gap, as evidenced by the comparisons in Figures~\ref{fig:real_image} and~\ref{fig:direct_output}, thereby substantially degrading the effectiveness of the physical attack.

Therefore, to improve the visual consistency between digital-domain attack optimization and physical-domain deployment, we introduce a lightweight MLP-based color calibration module before applying the homography transformation. \hzx{Specifically, we construct a reference color board, denoted as $\mathcal{B}_d$, and capture its printed counterpart, $\mathcal{B}_f$, under real-world conditions. The resulting dataset $\{\mathcal{B}_d, \mathcal{B}_f\}$ is used to train the MLP to learn the mapping between these spaces.} The architecture of this module is summarized in Table~\ref{tab:color_calibration_network}. As shown in Fig.~\ref{fig:color_calibration_comparison}, incorporating color calibration makes the synthesized image more closely resemble its real-world counterpart, thereby facilitating the physical realization of the attack.

\begin{table}[t]
    \centering
    \caption{Architecture of the color calibration network.}
    \label{tab:color_calibration_network}
    \begin{tabular}{lccc}
        \toprule
        \textbf{Layer} & \textbf{Type} & \textbf{Input Dim.} & \textbf{Output Dim.} \\
        \midrule
        1 & Linear     & 3  & 16 \\
        2 & LeakyReLU  & 16 & 16 \\
        3 & Linear     & 16 & 16 \\
        4 & LeakyReLU  & 16 & 16 \\
        5 & Linear     & 16 & 3 \\
        \bottomrule
    \end{tabular}
\end{table}

\subsection{Impact of TV Loss Weight}
We study the effect of the TV loss weight $\lambda_3$ on the stealthiness of the adversarial patches, with a particular focus on high-frequency artifacts under PGD optimization. As reported in Table~\ref{tab:tv_loss_weight}, increasing $\lambda_3$ consistently suppresses high-frequency components in the generated patches. Specifically, the high-frequency ratio is reduced by $98.7\%$ as $\lambda_3$ increases from $0.5$ to $100$. These results show that stronger TV regularization promotes smoother patches and reduces high-frequency artifacts, improving stealthiness. Moreover, we verify that the generated patches retain their attack effectiveness while maintaining high stealthiness.
\begin{table}[t]
\centering
\small
\caption{\textbf{Results of TV regularization on high-frequency artifacts.}
Relative drop is computed with respect to $\lambda_3=0.5$.}
\label{tab:tv_loss_weight}
\begin{tabular}{lcccc}
\toprule
TV Loss weight $\lambda_3$ & 0.5 & 1 & 10 & 100 \\
\midrule
High-frequency ratio $\downarrow$& 0.283 & 0.253 & 0.016 & 0.004 \\
Relative drop(\%) $\uparrow$& 0.0 & 10.6 & 94.3 & 98.7 \\
\bottomrule
\end{tabular}
\vspace{-10pt}
\end{table}

\subsection{Additional Analysis for Occlusion-free Deployment}
\label{appendix:further_print}
\begin{figure*}[t]
  \vskip 0.2in
  \begin{center}
    \centerline{\includegraphics[width=0.9\textwidth]{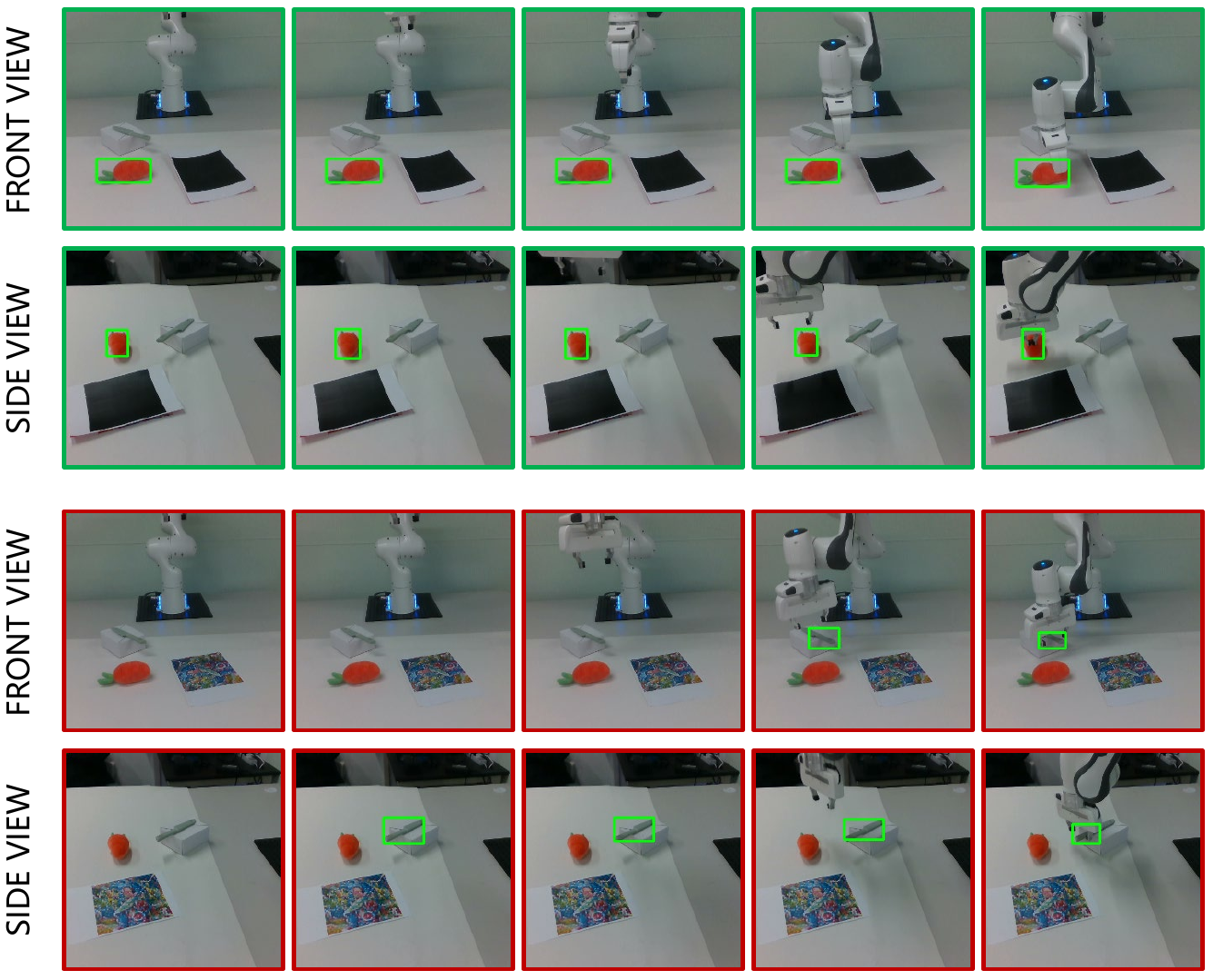}}
    \caption{
      Detailed occlusion-free deployment result.
    }
    \label{fig:full_real}
  \end{center}
    \vspace{-15pt}

\end{figure*}

Our real-world evaluation focuses on a hazardous redirection scenario, where the benign user instruction is ``pick up carrot'', but the adversary aims to hijack the VLA to ``pick up knife''. Across 15 independent trials, the attack successfully subverted the model's internal reasoning for at least one step in 13/15 (86.7\%) of the cases. Furthermore, in 5/15 (33.3\%) of the trials, the adversary achieved full behavior hijacking, successfully completing the malicious objective from start to finish. Fig.~\ref{fig:full_real} presents the synchronized multi-view observations (Front View and Side View) of this behavior hijacking process.

\begin{figure*}[t]
  \begin{center}
    \centerline{\includegraphics[width=0.6\textwidth]{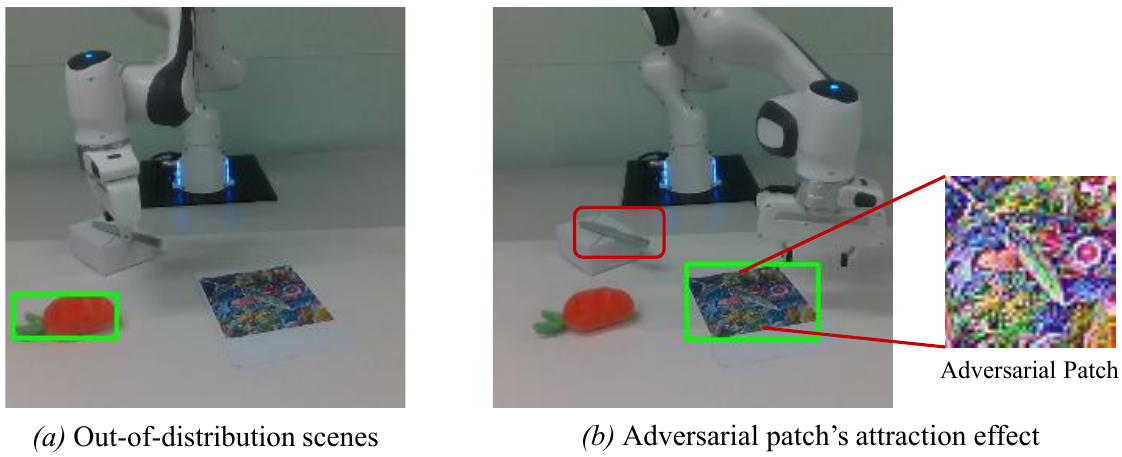}}
\caption{
\textbf{Illustration of representative failure cases.}
(a) Out-of-distribution observations reduce the transferability of the optimized adversarial patch, especially when large robot-arm motions introduce substantial visual changes.
(b) The adversarial patch may induce an attraction effect on CoT prediction, causing GraspVLA to localize the patch itself rather than the intended target object.
}
    \label{fig:failure_case}
  \end{center}
\end{figure*}

\textbf{Failure Cases Study.} We further examine the failure cases and identify two primary causes: out-of-distribution scenes and the attraction effect induced by the adversarial patch, as shown in Fig.~\ref{fig:failure_case}.

First, although the patch demonstrates a certain degree of scene-level generalization, its effectiveness degrades when the observed images differ substantially from those seen during optimization. This limitation becomes particularly pronounced under large robot-arm motions, which introduce significant visual changes into the observation. Moreover, for VLA models that use flow matching as the action expert, action generation is inherently multimodal due to the randomness of the initial noise. This additional stochasticity further increases the difficulty of reliably transferring the optimized patch across scenes and observations.

Second, we observe that the patch may exert a strong attraction effect on CoT prediction. In several cases, the bounding box predicted by GraspVLA for the target object incorrectly covers the patch itself rather than the intended object. One possible explanation is that, during optimization, the patch gradually acquires texture cues similar to those of the target object (\textit{e.g.}, the green knife), thereby confusing the VLA.

\subsection{Additional Analysis of Object-occluded Deployment}
\label{appendix:further_table}

We further evaluate \textbf{TRAP} under a more natural tablecloth-style deployment, where the optimized adversarial patch is enlarged and used as a tablecloth, with objects placed directly on top of it. Compared with the simulation-aligned printed-patch setting, this deployment is substantially more challenging: the patch is no longer an isolated planar pattern, but is partially occluded by task-relevant objects and distractors, while also being expected to remain visually inconspicuous in the scene.

\begin{figure*}[t]
  \centering

  \begin{subfigure}{0.9\textwidth}
    \centering
    \includegraphics[width=\linewidth]{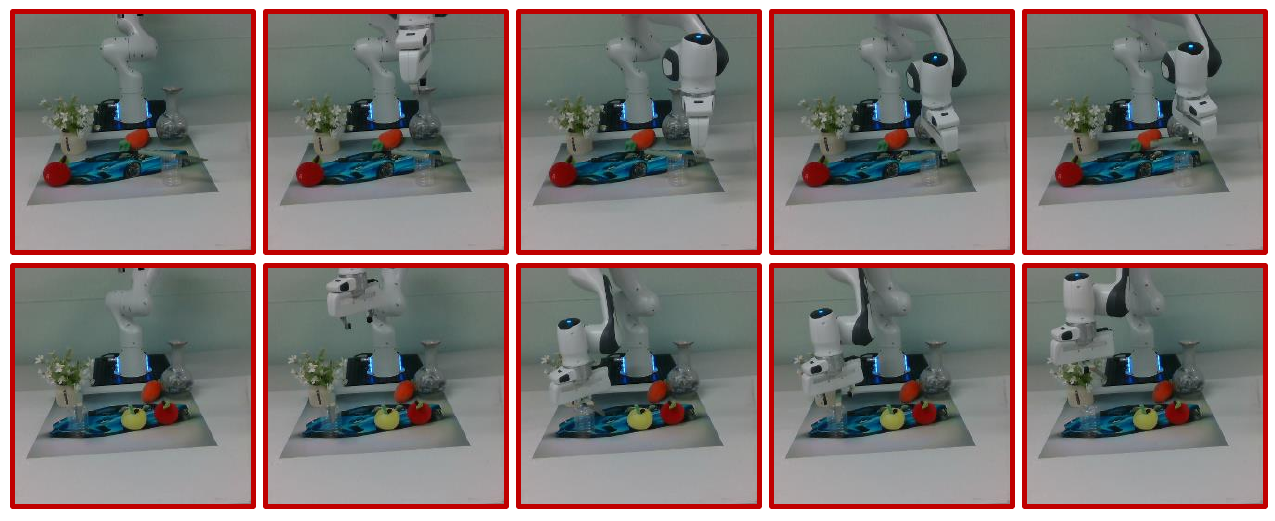}
    \caption{Evaluation under different layouts.}
    \label{fig:tablecloth_layout}
  \end{subfigure}

  \vspace{0.5em}

  \begin{subfigure}{0.9\textwidth}
    \centering
    \includegraphics[width=\linewidth]{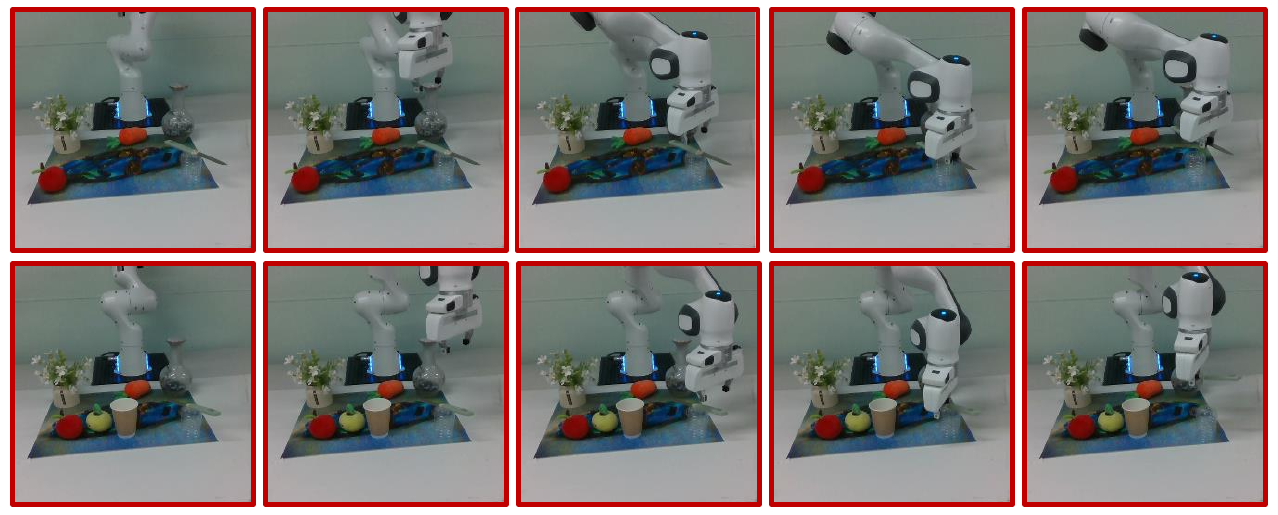}
    \caption{Evaluation under add/remove objects}
    \label{fig:tablecloth_distractor}
  \end{subfigure}

  \vspace{0.5em}

  \begin{subfigure}{0.9\textwidth}
    \centering
    \includegraphics[width=\linewidth]{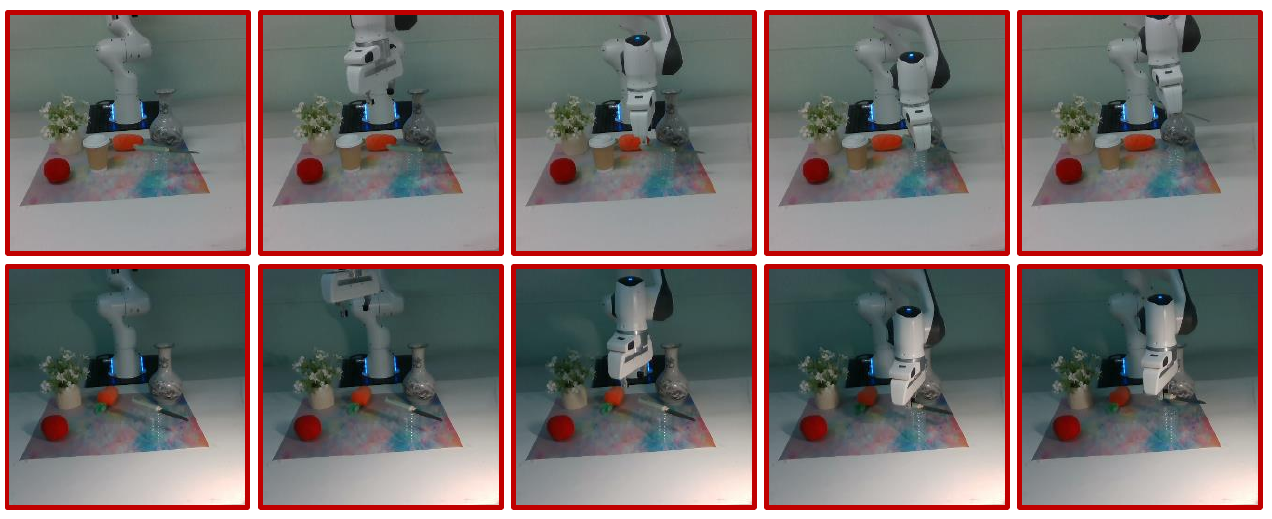}
    \caption{Evaluation under different lighting conditions.}
    \label{fig:tablecloth_lighting}
  \end{subfigure}

  \caption{
    Illustration of evaluation under different deployment variants.
  }
  \label{fig:deployment_variants}
\end{figure*}

The results show that \textbf{TRAP} remains effective under this more realistic physical setting. In particular, the tablecloth-style patch can still hijack the VLA behavior toward the adversarial target while preserving a semantically meaningful appearance. We further explore the robustness of the patch under several deployment variations, including different object layouts (Fig.~\ref{fig:tablecloth_layout}), additional distractor objects (Fig.~\ref{fig:tablecloth_distractor}), and different lighting conditions (Fig.~\ref{fig:tablecloth_lighting}). Across these variations, the patch consistently induces the target behavior, suggesting that the attack does not rely on a fixed tabletop configuration.

\begin{figure*}[t]
  \begin{center}
    \centerline{\includegraphics[width=0.8\textwidth]{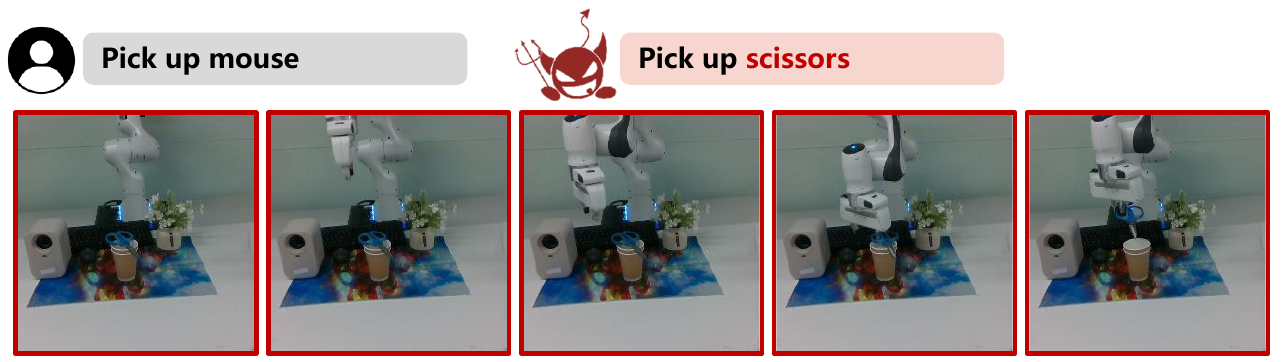}}
    \caption{
      \textbf{Illustration of another hazardous task.} The benign user instruction is ``pick up mouse'', while the adversary aims to hijack the robot toward ``pick up scissors''.
    }
    \label{fig:tablecloth_scissors}
  \end{center}
\end{figure*}

In addition to the ``carrot-to-knife" hazardous-redirection task studied in the main experiment, we further evaluate another harmful object-related scenario. In this case, the benign user instruction is ``pick up mouse'', while the adversary aims to hijack the robot toward ``pick up scissors''. As shown in Fig.~\ref{fig:tablecloth_scissors}, \textbf{TRAP} can also induce this harmful target behavior, further demonstrating its effectiveness in tablecloth-style physical deployments.


\end{document}